\documentclass[12pt]{article}
\usepackage{graphicx}
\usepackage{amsmath}
\usepackage{amsfonts}
\usepackage{amssymb}
\newtheorem{theorem}{Theorem}[section]
\newtheorem{definition}[theorem]{Definition}

\numberwithin{equation}{section}

\begin{document}

\title{Geometric Asymptotics and Beyond}

\author{
Helmut Friedrich\\ 
Max-Planck-Institut f\"ur Gravitationsphysik\\
Am M\"uhlenberg 1\\
14476 Golm, Germany}

\maketitle

{\footnotesize





\section{Introduction}

Enormous progress has been made in the last hundred years  in the investigation of  Einstein's field  equations, their solutions, their solution manifold, and thus  the content of the theory.
Research on the equations has been largely  guided by physical ideas but
unforeseen mathematical results repeatedly asked for  revisions, a   
process that may not have come to an  end yet \cite{friedrich:DPG}.
Mathematical analysis including  approximations and formal expansions led Einstein to an approximate notion of gravitational radiation and prepared the way to the invariant concept available today. Explicit solutions, mainly dealing with  highly symmetric and idealized situations, 
revealed  physical phenomena on global scales (horizons, black holes, singularities) 
which had  not been anticipated
and which  sometimes took years to be absorbed into a coherent world view. The abstract analysis of the field equations, dealing with the existence of solutions and their parametrization  in terms of boundary data,  evolved rather slowly, even after the breakthrough marked by the work of  Choquet-Bruhat \cite{choquet-bruhat:1952}. More recently, it became a tool to establish the existence and investigate the details of solutions with distinguished properties. Numerical techniques
allowed researchers to foray into domains  of the solution manifold hardly accessible by analytic methods  \cite{choptuik} and the calculation of quantitative data for comparisons with physical observations. 
In fact, certain data calculated for   astrophysical processes seem to have led to new observational 
results  \cite{komossa:2012}.
These activities went through layers of generalizations and refinements, discoveries of unexpected features, deepening of physical insights and ever more sophisticated  methods  of mathematical analysis.  In the following  this  process will be illustrated 
by a particular line of research. Originally motivated by 
the quest to `understand' gravitational radiation, it opened new views on the global space-time
structure

Because there exists now a vast literature on the subject I will have to focus in the following on  particular aspects of it. To keep the list of references at a reasonable length I shall often cite articles which appeared at  later stages in the development of some topic and give detailed refererence to earlier work. 
It should also be noted that the statements below which begin with a key word written in bold letters followed by some references may not always be found verbatim in these references  but follow immediately from the results and arguments given there.

\vspace{.1cm}

The second half of the 1950's  saw an intense activity concerned with attempts to find a covariant  concept of gravitational radiation  in non-linear general relativity  that had no need of approximation arguments. It is difficult to give a `correct' historical account of this process and the contributions of the various researchers. 
I just  point out a few highlights and refer to the literature for a more complete picture. 

Pirani  raises the question and tries to characterize radiative fields in terms of the Petrov structure of their curvature tensor  \cite{pirani:1957}. He observes that pointwise algebraic  considerations  hardly suffice and proposes to study the evolution of the Petrov structure by analyzing the  Bianchi identity $\nabla_{\mu}R^{\mu}\,_{\nu \lambda \rho} = 0$. 
Emphasizing the role of asymptotic  domains in radiation problems, Trautman brings a global aspect into the discussion  \cite{trautman:1958a}, \cite{trautman:1958c}. 
Defining coordinate systems $x ^{\mu}$ near {\it space-like infinity} that satisfy conditions analogous to Sommerfeld's `Ausstrahlungsbedingungen', he discusses on the space-like slices $x^0 = const.$ 
a total energy-momentum formula which may be considered as a precursor of a formula introduced
later at null infinity. Trying to integrate these considerations into a coherent picture, Sachs shifts the point of view and considers
 the field  along  {\it outgoing null geodesic congruences extending  to infinity}. Analysing  the Bianchi equations in a pseudo-orthonormal frame, he is led to suggest  that the curvature tensor shows at infinity a characteristic  {\it peeling-off behaviour}  of components related to the Petrov types \cite{sachs:waves VI}.

A decisive step is taken  when Bondi et al \cite{bondi:et.al}, Sachs \cite{sachs:waves VIII}, and Newman and Penrose \cite{newman:penrose} consider what in the language of today amounts to the 
 {\it asymptotic characteristic initial value problem}, where data are prescribed on an outgoing null hypersurface and at {\it future null infinity}.  They specify the asymptotic behaviour of the fields  in terms of distinguished null coordinates whose hypersurfaces  are  ruled by null geodesics extending to future null infinity. The behaviour of the fields is analyzed in terms of formal expansion in powers of $1/ r$ as $r \rightarrow \infty$,  where $r$ is a suitable parameter along the null geodesics.  In \cite{newman:penrose} the Bianchi equation is included into the system and the equations are expressed in terms of a pseudo-orthonormal  (more precisely, a spin) frame.
The required asymptotic behaviour is expressed in terms of the conformal Weyl tensor  and includes the peeling property. In  \cite{bondi:et.al} and \cite{sachs:waves VIII} are considered mass terms and mass-loss relations which are interpreted as indication that the mass must  decrease if the system is radiating. This result strongly supports the view of being  on the right track.

This development reached an apex when Penrose  introduces  an idea
that makes the role of the conformal structure, which features in the analysis above 
in terms of null hypersurfaces and null geodesics, quite explicit and offers at the same time a new view on the asymptotic and, in particular, on the global structure of gravitational  fields \cite{penrose:1963}. Let $( \hat{M}, \hat{g})$ denote the space-time of a 
{\it self-gravitating isolated system} that is so far away from other such systems that one can essentially ignore the influence of the latter, possibly with the exception of their gravitational radiation effects. Penrose proposes that the asymptotic behaviour of $( \hat{M}, \hat{g})$ is characterized by the following property:

\begin{definition}
\label{conf-ext}
The space-time  $( \hat{M}, \hat{g})$ is said to be asymptotically simple if there exists a space-time $(M, g)$ with boundary ${\cal J} \neq \emptyset$ such that $\hat{M}$ 
can be diffeomorphically identified with the interior $\hat{M} = M \setminus {\cal J}$ of $M$ so that

\vspace{.1cm}

\hspace*{5cm}$g_{\mu\nu} = \Omega^2\,\hat{g}_{\mu\nu} \quad \mbox{on} \quad \hat{M}$,

\vspace{.1cm}

\noindent
with a conformal factor  $\Omega$ which is a boundary defining function on $M$ that satisfies

\vspace{.1cm}

\hspace*{3cm}$\Omega > 0  \quad \mbox{on} \quad \hat{M}, \quad \Omega = \,0, \,\,d\Omega \neq 0
\quad \mbox{on} \quad {\cal J}$.

\end{definition}

The requirements only concern the asymptotic properties of the conformal structure of $\hat{g}$ since they are invariant under rescalings of the form 
$(\Omega, \hat{g}) \rightarrow (\theta\,\Omega, \theta^2\,\hat{g})$ with positive functions $\theta$. Conditions which restrict this freedom will be referred to as {\it conformal gauge conditions}.
Usually there are also stated  conditions which ensure that the {\it conformal boundary} ${\cal J}$ 
is as complete as possible. This is also related to the question of the uniqueness of the conformal boundary.
We shall not discuss these questions here. In the situations considered below, the conformal boundary will be uniquely determined by the evolution process defined  by the field equations once suitable initial data are prescribed and be it will  independent to the conformal gauge employed. 

The degree of  smoothness with which the conditions of the definition can be achieved  reflects in a  precise way the fall-off behaviour of the physical fields. It will be seen  that it can pose some of the most delicate problems. Unless remarked to the contrary, however, it will be supposed that $\Omega$ and $g$  (and the functions $\theta$ above) are of class $C^{\infty}$ on $M$.

It follows then that $\hat{g}$-null geodesics are complete in those directions in which they approach 
${\cal J}$. The set ${\cal J}$ thus represents a {\it null infinity} for the `physical' space-time $( \hat{M}, \hat{g})$, its points being endpoints of $\hat{g}$-null geodesics.
In general one would add conditions which ensure that no essential boundary points are left out in the construction of $M$ and for some types of systems
 there would be derived consequences such as a splitting 
${\cal J} = {\cal J}^- \cup \,{\cal J}^+$ of the boundary into two components ${\cal J}^{\pm}$ which represent the past/future endpoints and thus past/future null infinity  respectively. This will be seen later in concrete situations.
The impression that the definition disburdens us from considering distinguished coordinates is not quite correct.
To decide whether a given space-time is asymptotically simple one has to extend its differentiable structure, which amounts to singling out (if possible) coordinate systems which can be extended consistent with the conditions above.

\vspace{.3cm}

Statements about asymptotic simplicity gain significance  if  Einstein's field equations are involved. 
Penrose extended the range of application of Definition \ref{conf-ext} beyond that of isolated systems and also considered solutions with non-vanishing cosmological constant 
$\lambda$ \cite{penrose:1965}. 
In general a smooth solution $( \hat{M}, \hat{g})$ to  the vacuum equations  with cosmological constant $\lambda$ does not even satisfy the conditions with 
an extension $(M, g, \Omega)$ of finite differentiability.
But if it does, with sufficient smoothness, it  follows that  the field equations determine the  causal nature of the boundary. The set 
${\cal J}$ will be time-like, null, or space-like depending on $\lambda$ being negative, zero, or positive (assuming the signature $(-, +, +, +)$). While this is a simple consequence of the field equations,
 it shows that the sign of the cosmological constant has far-ranging effects on the overall behaviour of the solutions.
Today this is taken for granted, but at a time when local coordinates still dominated the way of looking at  space-times, it must have come as a revelation.

The conformal Weyl tensor $C^{\mu}\,_{\nu \kappa \sigma}[ \hat{g}]$ of $ \hat{g}$ goes to zero at ${\cal J}$ if there exists a conformal extension $(M, g, \Omega)$ of sufficiently high differentiability \cite{penrose:1965}. 
In fact, observing that $C^{\mu}\,_{\nu \kappa \sigma}[ \hat{g}]  = C^{\mu}\,_{\nu \kappa \sigma}[g]$ on 
$\hat{M}$,
one can easily specify precise smoothness conditions on $(M, g, \Omega)$ which imply that 
$\nabla_{\mu}\Omega\,C^{\mu}\,_{\nu \kappa \sigma}[g] = 0$ on ${\cal J}$. 
If $\lambda \neq 0$, so that $g(d\Omega, d\Omega) \neq 0$,
 pointwise algebra shows that $C^{\mu}\,_{\nu \kappa \sigma}[g] = 0$ on ${\cal J}$. If $\lambda = 0$ the conclusion is more subtle. The argument given in  \cite{penrose:1965} contains an implicit assumption on the smoothness of the conformal extension $(M, g, \Omega)$ 
which requires difficult global considerations for its justification (in the review of the argument given in \cite{friedrich:tueb} this assumption is taken as the starting point). We shall see below, that the situation is not completely understood yet. Therefore it will sometimes be convenient to use the notation 
 $ {\cal J}_*$ to refer to  
null infinity  simply as a unstructured set of fictitious endpoints of null geodesics.

Suppose there does exist a smooth conformal extension satisfying  the requirements of Definition \ref{conf-ext}.
Then many questions  concerning the asymptotic structure  which required delicate limits before can conveniently  be studied in terms of local differential geometry. The conditions then comprise 
fall-off properties  of the curvature tensor which are such that  
$W^{\mu}\,_{\nu \kappa \sigma}  = \Omega^{-1}\,C^{\mu}\,_{\nu \kappa \sigma}[ \hat{g}] $ extends smoothly to the boundary ${\cal J}$ and it follows by pure algebra that the Sachs peeling conditions are
satisfied  \cite{penrose:1974}. In the vacuum case with $\lambda = 0$ the setting provides a precise notion of {\it radiation field}. It is given on ${\cal J}^+$ (say)
by  the complex-valued function
$\psi_0 = \psi_{ABCD}\,\iota^A\,\iota^B\,\iota^C\,\iota^C$ where  $\psi_{ABCD}$ is the symmetric spinor field which represents the limit of the tensor $W^{\mu}\,_{\nu \kappa \sigma}$ to ${\cal J}^+$ and
$\iota^A$ is a two-index spinor field so that $\iota^A\,\bar{\iota}^{A'}$ is tangential to the null generators of the null hypersurface ${\cal J}^+$. For a discussion of the relations between the radiation field, the mass, and the mass loss along ${\cal J}^+$ we refer to \cite{penrose:1974}.

The situation described in Definition \ref{conf-ext} will be referred to by saying that {\it $(\hat{M}, \hat{g})$ admits a smooth conformal extension at null infinity}. We emphasize that this comprises the smooth extendibility of the metric $g$ as well as that of the conformal factor $\Omega$ with the required properties.  In the context of the Einstein equations both requirements are important, because they relate in a precise way to the decay behaviour of the gravitational  field indicated above. 
That these things should not be taken for granted  is illustrated by the following example. The space-time with manifold $\hat{M} = \mathbb{R}^2 \times \mathbb{S}^2$ and metric $\hat {g} = - dt^2 + dr^2 + \cosh^2r\,h_{\mathbb{S}^2}$ is geodesically complete. That it admits a conformal extension is seen as follows \cite{bizon:mach:2014}. With the transformation defined by
$\cosh \tau \,\cos \psi = \tanh r$, $\cosh \tau \,\sin \psi = \cosh t \,\cosh^{-1} r$, 
$\sinh \tau = \sinh t \,\cosh^{-1} r$ and the  conformal factor $\Omega = \cosh^{-1} r$ one obtains
$\Omega^2\,\hat{g} = g \equiv - d\tau^2 + \cosh^2 \tau\,d\psi^2 + h_{\mathbb{S}^2}$. The metric $g$ is the Nariai solution which lives in fact on $M = \mathbb{R} \times   \mathbb{S}^1  \times  \mathbb{S}^2$. The conformal embedding of $(\hat{M}, \hat {g} )$  into $(M, g)$ so obtained covers only the domain where 
$|\cosh \tau\, \cos \psi| < 1$, $0 < \psi < \pi$. This domain is bounded by null hypersurfaces on which
$|\cosh \tau\, \cos \psi| = 1$. These can be understood as defining a conformal boundary for the space-time 
$(\hat{M}, \hat {g} )$. While the metric $\Omega^2\,\hat{g} $ extends smoothly to this boundary, the function $\Omega$, which can be written $\Omega = \sqrt{1 - \cosh^2 \tau\, \cos^2 \psi}$, only  extends continuously with $\Omega = 0$ but $d\Omega$ divergent on the boundary. Moreover, the conformal Weyl tensor $C^{\mu}\,_{\nu \lambda \rho}[g]$  does not go to zero  on this boundary.

\section{Field equations and conformal rescalings.}

The early studies of the notion of asymptotic simplicity focus on geometrical and physical issues, the existence  of solutions satisfying the conditions is discussed mainly in terms of  explicit, usually static or stationary ones, which have vanishing radiation field \cite{Geroch: 1977}, \cite{hawking:ellis:1973}, \cite{penrose:1965}. There have been found  explicit vacuum solutions which admit pieces of a smooth conformal boundary with non-vanishing radiation field \cite{bicak}, but non of them arises from smooth, asymptotically flat, geodesically complete data on a Cauchy hypersurface and admits a conformal boundary that satisfies reasonable completeness conditions
\cite{Geroch-Horowitz:1978}.
At the time, anyone toiling at the abstract analysis of the Cauchy problem for the non-linear Einstein equations must have registered with surprise that the complicated global problem of characterizing the asymptotic behaviour of solutions 
should allow, in any generality, an answer in so simple and clean geometric  terms as used in 
Definition \ref{conf-ext}.  
In fact, it has been argued  from early on 
that the assumptions on the asymptotic behaviour underlying  the work referred to above
might be too stringent and that consistent formal expansions at null infinity can also obtained under more general assumptions \cite{chrusciel-maccallum-singleton:1995}, \cite{couch-torrence:1972}.

Many questions remained open and we shall address here only a few of them. Physically a particularly interesting one, which will obviously become most important once gravitational radiation can be measured directly, 
refers to fields of isolated self-gravitating systems:
 {\it  What kind of information on the structure of the sources can be extracted from the radiation field ?}
This question does not only ask for  qualitative control on the  far fields and the (massive)
sources, but also for quantitative results. These can, in particular in the case of strong and highly dynamical fields, only be obtained by numerical methods. 
In this article  we shall mainly be interested in a more basic problem:  {\it How rich is the class of solutions satisfying Definition \ref{conf-ext} ?} Besides providing mathematical results on the structure of the field equations and the large scale nature  of the solutions (which should also help with the first question above)
the complete answer should include a discussion of the  physical significance 
of  possible obstructions to asymptotic simplicity. 
We shall be interested  in the following  in solutions of four space-time dimensions which are general in the sense that they are not required to  have symmetries.  
 
 Answers to the second  question above can only be obtained by methods of general abstract analysis and global or semi-global existence results. Provided the extension is smooth,  asymptotic simplicity makes a  statement about the asymptotic behaviour that is absolutely sharp. 
 This makes the concept  most delicate from the PDE point of view and harbours the danger of leading to
undesired restrictions. On the other hand, it may give deeper insight into the mathematical structure and the physical meaning of solutions. In the following sections we will have various opportunities to discuss this dichotomy. 

In the early 1970's there existed a huge gap between the formal expansion type studies at null infinity and the abstract existence theory for Einstein's field equations. At the time, the latter supplied  existence results local in time for the wave equation obtained from Einstein's equations in harmonic coordinates. The two approaches to Einstein's field equations were completely unrelated.  An effort to understand the field equations in the conformal  setting combined with a search for alternative  ways to exploit  the intrinsic hyperbolicity of the Einstein equations showed, however, that Penrose's emphasis on the conformal structure might lead to new methods for the existence problem \cite{friedrich:1981a}.

 If the vacuum equation  $R_{\mu \nu}[\hat{g}] = 0$ for the `physical' metric $\hat{g}_{\mu \nu}$ is expressed in terms of a conformal factor $\Omega$ and the  conformal metric 
 $g_{\mu \nu} = \Omega^2\,\hat{g}_{\mu \nu}$,  it reads
 \begin{equation}
 \label{Ric[g]}
R_{\mu \nu}[g] = - 2\,\Omega^{-1}\,\nabla_{\mu}\nabla_{\nu}\Omega
- g_{\mu\nu}\,g^{\alpha \beta}\left(
\Omega^{-1}\,\nabla_{\alpha}\nabla_{\beta}\Omega
 - 3\,\Omega^{-2}\,\nabla_{\alpha}\Omega\,\nabla_{\beta}\Omega \right).
\end{equation}
The right hand side  becomes singular where $\Omega \rightarrow 0$. It is one of the remarkable features of the Einstein equations  that (\ref{Ric[g]}) can be included into a larger system of equations, the {\it conformal field equations} discussed below , which avoids this type of singularity. By itself this might not help much
but it turns out that the resulting system admits hyperbolic reductions \cite{friedrich:asymp-sym-hyp:1981b}. It generalizes the Einstein equations by being  equivalent to them where $\Omega \neq 0$ but still implying hyperbolic evolution equations where $\Omega = 0$. 

It is clear that  in controlling  the large scale behaviour of their solutions  in terms of estimates on the physical field $\hat{g}$ and fields derived from it,  the conformal properties of the equations exhibited here must play implicitly an important role. However, the explicit use of these properties in terms of
the conformal field equations seemed to offer  ways  to calculate 
global or semi-global solutions to Einstein's field equations numerically on finite grids.
Formulating corresponding initial or initial-boundary value problems and working out their 
qualitative consequences for the global space-time structure analytically
therefore suggested itself as a way to also obtain quantitative results and thus answers to the first question asked above as well.

\subsection{Conformal field equations.}

Einstein's equations  with cosmological constant $\lambda$ and energy-momentum tensor 
$\hat{T}_{\mu\nu}$,
\begin{equation}
\label{Einst-equ}
\hat{R}_{\mu\nu} - \frac{1}{2}\,\hat{R}\,\hat{g}_{\mu\nu} + \lambda\,\hat{g}_{\mu\nu}
= \kappa\,\hat{T}_{\mu\nu},
\end{equation}
coupled to suitable matter field equations, are reexpressed
in terms of a conformal factor $\Omega$, the metric $g_{\mu\nu} = \Omega^2\,\hat{g}_{\mu\nu}$, 
suitably  transformed matter fields, and the derived fields
 
\vspace{.1cm}

\hspace*{.1cm}$s \equiv \frac{1}{4}\,\nabla_{\rho}\,\nabla^{\rho}\Omega + \frac{1}{24}\,\Omega\,R, 
\quad \,\, 
L_{\mu \nu} = \frac{1}{2}\left(R_{\mu \nu} - \frac{1}{6}\,R\,g_{\mu \nu}\right),
\quad \,\, 
W^{\mu}\,_{\rho \nu \lambda}  \equiv \Omega^{-1}\,C^{\mu}\,_{\rho \nu \lambda}$, 

\vspace{.1cm}

\hspace*{2cm}$T^*_{\rho \mu} 
\equiv \hat{T}_{\rho \mu} - \frac{1}{4}\,\hat{T}\,\hat{g}_{\rho \mu}, 
\quad \quad 
\hat{\nabla}_{\rho} \hat{L}_{\mu \nu} \equiv \frac{\kappa}{2}\,\hat{\nabla}_{\rho} \left(\hat{T}_{\mu \nu} 
- \frac{1}{3}\,\hat{T}\,\hat{g}_{\mu \nu} 
\right)$,

\vspace{.1cm}

\noindent
where $\hat{\nabla}_{\mu}$, $\nabla_{\mu}$ and $\hat{C}^{\mu}\,_{\rho \nu \lambda}$, 
$C^{\mu}\,_{\rho \nu \lambda}$ denote the covariant derivative operators and the 
conformal Weyl tensors of the metrics  
$\hat{g}$ and $g$ respectively, $L_{\mu \nu}$  is the Schouten tensor of $g$,
and $\hat{T}$ the $\hat{g}$-trace of $\hat{T}_{\mu\nu}$.
Equations (\ref{Einst-equ})  imply the {\it conformal field equations} (\cite{friedrich:1981a}, \cite{friedrich:asymp-sym-hyp:1981b}, \cite{friedrich:1991})
\begin{equation}
\label{mat-B-Ric-scal-transf}
6\,\Omega\,s - 3\,\nabla_{\rho}\Omega\,\nabla^{\rho}\Omega - 
\lambda = - \frac{\kappa}{4}\,\hat{T},
\end{equation}
\begin{equation}
\label{mat-B-Ric-ten-transf}
\nabla_{\mu}\,\nabla_{\nu}\Omega + \,\Omega\,L_{\mu\nu} - s\,g_{\mu\nu}
= \frac{\kappa}{2}\,\Omega\,T^*_{\mu\nu},
\end{equation}
\begin{equation}
\label{mat-s-equ}
\nabla_{\mu}\,s + \nabla^{\rho}\Omega\,L_{\rho\mu} 
= \frac{\kappa}{2}\,\nabla^{\rho}\Omega\,T^*_{\rho \mu}
- \frac{\kappa}{24\,\Omega}\,\hat{\nabla}_{\mu}\,\hat{T},
\end{equation}
\begin{equation}
\label{mat-L-equ}
\nabla_{\nu}\,L_{\lambda \rho} 
- \nabla_{\lambda}\,L_{\nu \rho} - 
\nabla_{\mu}\Omega\,\,W^{\mu}\,_{\rho \nu \lambda} 
= 2\,\hat{\nabla}_{[\nu}\,\hat{L}_{\lambda] \rho},
\end{equation}
\begin{equation}
\label{mat-W-equ}
\nabla_{\mu}\,W^{\mu}\,_{\rho \nu \lambda} 
= \frac{2}{\Omega}\,\hat{\nabla}_{[\nu}\,\hat{L}_{\lambda] \rho},
\end{equation}
where all contractions are performed with the metric $g$. The first two equations are just a rewrite of
(\ref{Einst-equ}). They imply the differential identity  (\ref{mat-s-equ}). Equation (\ref{mat-W-equ}) is obtained from the contracted Bianchi identity for $\hat{g}$ with the conformally covariant relations

\vspace{.1cm}

\hspace*{2cm}$C^{\mu}\,_{\rho \nu \lambda} = \hat{C}^{\mu}\,_{\rho \nu \lambda},
\quad \quad
\nabla_{\mu}(\Omega^{-1}\,C^{\mu}\,_{\rho \nu \lambda} ) = 
\Omega^{-1}\,\hat{\nabla}_{\mu}\,\hat{C}^{\mu}\,_{\rho \nu \lambda}$.

\vspace{.1cm}

\noindent
With (\ref{mat-W-equ}) equation (\ref{mat-L-equ}) is obtained from the Bianchi identity
for $g$.
These equations must be supplemented  by equations which relate the tensorial unknowns
\begin{equation}
\label{tensorial-unknowns}
\Omega, \quad s, \quad L_{\mu \nu}, \quad W^{\mu}\,_{\rho \nu \lambda},
\end{equation}
to the metric  $g$ and the connection $\nabla$. One possibility to do this is by introducing 
a $g$-orthonormal frame field $\{e_k \}_{k = 0, \ldots 3}$ and suitable coordinates $x^{\mu}$.
In terms of the frame coefficients $e^{\mu}\,_k$ and the 
connection coefficients $\Gamma_i\,^k\,_j$  defined by the relations 
$e_k = e^{\mu}\,_k\,\partial_{x^{\mu}}$ and 
$\nabla_i\,e_j = \Gamma_i\,^k\,_j\,e_k$ so that
$g(e_i, e_j) 
= g_{\mu\nu}\,e^{\mu}\,_i\,e^{\nu}\,_j  = \eta_{ij}$ and 
$\Gamma_{ijk} = - \Gamma_{ikj}$,
where $\nabla_i = \nabla_{e_i}$ and $\Gamma_{ijk} = \Gamma_i\,^l\,_k\,\eta_{lj}$,  
the {\it structural equations}  then  take the form of the
{\it torsion-free condition} 
\begin{equation}
\label{torsion-free condition}
e^{\mu}\,_{i,\,\nu}\,e^{\nu}\,_{j}
 - e^{\mu}\,_{j,\,\nu}\,e^{\nu}\,_{i} 
= (\Gamma_{j}\,^{k}\,_{i} - \Gamma_{i}\,^{k}\,_{j})\,e^{\mu}\,_{k},
\end{equation}
and the {\it Ricci identity}
\begin{equation}
\label{Ricci identity}
\Gamma_l\,^i\,_{j,\,\mu}\,e^{\mu}\,_k - \Gamma_k\,^i\,_{j,\,\mu}\,e^{\mu}\,_l
+ 2\,\Gamma_{[k}\,^{i\,p}\,\Gamma_{l]pj}
- 2\,\Gamma_{[k}\,^p\,_{l]}\,\Gamma_p\,^i\,_j
\end{equation}
\[
= \Omega\,W^i\,_{jkl}
+ 2\,\{g^i\,_{[k}\,L_{l] j} + L^i\,_{ [k}\,g_{l] j}\}.
\]
If equations 
(\ref{mat-B-Ric-scal-transf}) to (\ref{mat-W-equ}) are expressed in terms of the frame and combined with the structural equations, they are equivalent  (ignoring subtleties which may arise in the case of very low differentiability) to Einstein's vacuum equations where
$\Omega > 0$ and in fact also where $\Omega < 0$ 
(with the replacement $(\Omega, s,  W_{ijkl}) \rightarrow (- \Omega, - s, - W_{ijkl})$ and the other fields unchanged the equations remain  satisfied as they stand).

If (\ref{mat-B-Ric-scal-transf}) holds on an initial slice, which can always be arranged, it will be 
satisfied as a consequence of the other equations.
In the {\it vacuum case} $\hat{T}_{\mu\nu} = 0$ the  right hand sides of equations 
(\ref{mat-B-Ric-ten-transf}) to (\ref{mat-W-equ}) vanish and all factors $1/\Omega$ in the equations are gone.  If  the Ricci scalar $R$ is prescribed as a function of the coordinates (which can locally be done in an arbitrary way) and the system (\ref{mat-B-Ric-ten-transf}) to (\ref{Ricci identity})
 is written with respect to a suitable choice of coordinates and frame field, it implies {\it reduced equations} which are {\it hyperbolic even where $\Omega$ changes sign}. Moreover, the evolution by the reduced system preserves the constraints and the gauge conditions.
 Various versions of hyperbolic systems are given in \cite{friedrich:asymp-sym-hyp:1981b}, 
\cite{friedrich:1986b}, \cite{friedrich:1991}, \cite{friedrich:tueb}.
 The case of zero rest-mass fields, for which the energy momentum tensor is trace free and the matter fields admit regular conformal representations is similar. 
For other matter models the situation is more difficult and will in general  result in equations where
the  $1/\Omega$ terms  cannot be removed. A remarkable case of non-zero rest-mass fields 
with non-vanishing trace of the energy momentum tensor where this can be done  will be discussed in section \ref{positive lambda}.

While the search for hyperbolic reductions which could be adapted to various particular  situations led to quite general reduction procedures,  in fields of isolated systems a precise analysis of the equations remained extremely difficult in the neighbourhood of space-like infinity.
Admitting besides conformal rescalings also transitions to Weyl connections, 
one obtains, however,  an extended version of the conformal field equations.
It  allows one to employ a type of  {\it geometric gauge}, based on {\it conformal geodesics} which has not been used in the context of the Einstein equations before. Apart from some freedom on the initial slice, the gauge  is then defined completely in terms of the conformal structure \cite{friedrich:AdS}, \cite{friedrich:2hyp red}. The resulting equations are referred to as the {\it general conformal field equations}.
The  existence results discussed in the following sections have been obtained with the conformal field equations above or such generalizations.

In the following years were studied modifications of the equations above, in particular systems of wave equations obtained by taking further derivatives \cite{choquet-bruhat:novello:1987}, \cite{paetz:2013}, and versions of  conformal field equations that apply to other situations. 
In  the asymptotically flat case the static and stationary vacuum  Einstein equations with $\lambda = 0$ were  shown to admit  conformal representations on the  3-dimensional quotient space which imply that the fields extend as real analytic fields to spatial infinity \cite{beig:simon1980}, \cite{beig:simon1981},  \cite{kundu:1981}. 
Tod discusses  big bang like {\it isotropic cosmological singularities}  that are 
characterized by the existence of conformal rescalings which, in contrast to the asymptotically simple case, {\it blow up} a neighborhood of the singularity so as to represent the big bang by a hypersurface   smoothly attached to the original space-time \cite{tod:1987}. These solutions were studied in the following by versions of conformal field equations which  include suitable matter fields and deal with some remaining singular terms \cite{RPAC-Newman: 1993},  \cite{tod:2003}, \cite{tod:2007}.

Though  the conformal field equations above do admit regular generalizations to higher dimensions,  the space-time dimension four is special for them \cite{friedrich:tueb}. Only in that dimension do they supply  hyperbolic evolutions systems of first order in the unknowns (\ref{tensorial-unknowns}). This raises the question whether there exist generalizations.
Various authors studied formal expansion at the (space- or time-like) conformal boundary ${\cal J}$ of  solutions to Einstein's equations with or without matter fields and with cosmological constant $\lambda \neq 0$ 
in space-time dimensions $n \ge 4$ \cite{graham:hirachi:2005}, \cite{rendall:2004}, \cite{starobinsky:1983}. In odd space-time dimensions   these expansions must include logarithmic terms to exploit the full freedom to prescribe data on ${\cal J}$.  The logarithmic terms not to occur on the {\it even dimensional } boundary ${\cal J }$ if the  {\it Fefferman-Graham obstruction tensor} ${\cal O}_{jk}$ of the free data vanishes on   ${\cal J }$. This tensor, 
defined by an operator of order $n$ for any metric $g$ on a manifold of even dimension $n \ge 4$, is 
trace-free, symmetric, conformally invariant,  and vanishes for metrics conformal to Einstein metrics \cite{graham:hirachi:2005}. 
Anderson uses the equation  ${\cal O}_{jk}[g] = 0$ as a conformal vacuum Einstein equation  \cite{anderson:2005a}.

\section{De Sitter-type solutions.}
\label{positive lambda}

In recent years  various observations suggested an accelerating expansion of the universe and all of them seem to be consistent with the assumption of a positive cosmological constant in Einstein's field equations \cite{davis:2014}. There exist  neither convincing theoretical explanations for the acceleration nor for the cosmological constant. Therefore it is of particular interest to understand 
the manifold of solutions to Einstein's field equations with positive $\lambda$.
In the vacuum case $\hat{T}_{\mu\nu} = 0$ we write equation (\ref{Einst-equ}) in the form 
\begin{equation}
\label{vac-lambda}
Ric[\hat{g}] = \lambda \hat{g}.
\end{equation}
The simply connected, conformally flat prototype  solutions to (\ref{vac-lambda}) with $\lambda = 3$ is 
 {\it de Sitter space}  $(\hat{M} = \mathbb{R} \times \mathbb{S}^{3}, 
 \hat{g} = - dt^2 + \cosh^2t\,\,h_{\, \mathbb{S}^{3}})$  where $h_{\, \mathbb{S}^{n}}$ denotes the standard metric on the unit $n$-sphere $ \mathbb{S}^{n}$.
With the coordinate transformation  $ t \rightarrow \tau = 2\,\arctan e^t - \frac{\pi}{2}$
and the conformal factor $\Omega = \cosh^{-1} t = \cos \tau $ one obtains the conformal representation
$\hat{M} = \,]- \frac{\pi}{2},  \frac{\pi}{2}[ \, \times \,\mathbb{S}^3$,
$g = \Omega^2\,\hat{g} =  - d\tau^2 + h_{\,\mathbb{S}^{3}}$.
The metric on the right hand side and the conformal factor extend smoothly to the manifold 
with boundary $M = \,[- \frac{\pi}{2},  \frac{\pi}{2}]  \, \times \,\mathbb{S}^3$. Their extensions will be denoted again by $g$ and $\Omega$. The  conformal boundary ${\cal J}$ splits into the  
components ${\cal J}^{\pm} =  \{\pm \,\frac{\pi}{2},\} \times \,\mathbb{S}^{3}\,$
which are space-like for $g$ and represent for $\hat{g}$ future and past null and time-like infinity respectively.

In the following we shall be interested in globally hyperbolic generalizations of this solution which either admit a smooth conformal extension with conformal boundary ${\cal J}^+$  in the future (which maps onto a ${\cal J}^-$ under time reversal), or conformal extensions in the past as well as in the future, with conformal boundaries ${\cal J}^{\pm}$. These solutions will be such that each null geodesics acquires precisely one future endpoint on ${\cal J}^+$ in the first case and 
precisely one past endpoint on ${\cal J}^-$ and one future end point on ${\cal J}^+$ in the second case. The hypersurfaces ${\cal J}^+$, resp. ${\cal J}^-$ and ${\cal J}^+$ will constitute Cauchy hypersurfaces for the conformal extension  $(M, g)$ and initial hypersurfaces suitable for the conformal field equations. 
To avoid misunderstandings arising from the fact  that the sign of the cosmological constant only becomes meaningful if the signature of the metric is fixed,
we shall refer to such space-times as {\it de Sitter type space-times}.

\subsection{Existence and stability results.}

To construct such solutions, suitable initial data are needed. Because there seem to be no natural boundary conditions for solutions with $\lambda > 0$, 
we shall assume in both types of initial data sets discussed below that {\it the $3$-manifold $S$ is compact}.
Moreover, only smooth initial data will be considered.
Let $\lambda$ denote a fixed positive number.

A {\it standard  initial data set for Einstein's vacuum field equations (\ref{vac-lambda})
with cosmological constant $\lambda$} consists of an orientable 3-dimensional Riemannian manifold 
$(S, \hat{h}_{ab})$ and a symmetric tensor field $\hat{\chi}_{ab}$ which satisfy on $S$ the Hamiltonian constraint
$R[\hat{h}] = \hat{\psi}_{ab}\,\hat{\psi}^{ab} + 2\,\lambda - \frac{2}{3}\,\hat{\chi}^2$
 and the momentum constraint
$\hat{D}^a\,\hat{\psi}_{ab} = \frac{2}{3}\,\hat{D}_b\hat{\chi}$,
where $\hat{D}$ and $R[\hat{h}] $ denote the Levi-Civita operator and the Ricci scalar of $\hat{h}$ and
we use the decomposition 
$\hat{\chi}_{ab} = \hat{\psi}_{ab} + \frac{1}{3}\,\hat{\chi}\,h_{ab}$ into $\hat{h}$-trace-free part and trace.

The set of such data will be denoted by $\cal D_S$. Any data set    
$(S, \hat{h}_{ab}, \hat{\chi}_{ab})$ in $\cal D_S$ determines a unique (up to diffeomeorphisms) maximal, globally  hyperbolic solution $(\hat{M}, \hat{g})$ to (\ref{vac-lambda}) which contains  a Cauchy hypersurface
$\hat{S}$ that is,  together with the first and second fundamental form induced on it by $\hat{g}$, diffeomorphic$(S, \hat{h}_{ab}, \hat{\chi}_{ab})$  \cite{choquet-bruhat:geroch:1969}.
We note that in the case $\hat{\chi}^2 < 3\,\lambda$ the  Hamiltonian constraint implies
$R[\hat{h}] > 0$ and thus a restriction on the conformal structure of $(S, \hat{h})$.
From the  initial data  $(\lambda, S, \hat{h}_{ab}, \hat{\chi}_{ab})$ can be derived  initial data  for the conformal field equations.

An {\it asymptotic  initial data set for Einstein's vacuum field equations (\ref{vac-lambda})
with cosmological constant $\lambda > 0$} consists of an orientable 3-dimensional Riemannian manifold $(S, h_{ab})$ and a symmetric tensor field $w_{ab}$ on $S$ that satisfies 
$w_a\,^a = 0$ and $D^a w_{ab} = 0$, where $D$ denotes the Levi-Civita operator of $h$.

\vspace{.1cm}

\noindent
{\bf Semi-global existence} \cite{friedrich:1986a}: {\it  An
asymptotic  initial data set $(\lambda, S, h_{ab}, w_{ab})$ for the vacuum field equations  
(\ref{vac-lambda})
determines a  unique, maximal, globally 
hyperbolic solution
$(\hat{M}, \,\hat{g})$ to $Ric[\hat{g}] = \lambda \hat{g}$
with $\hat{M} \sim \mathbb{R} \times S$
which admits a smooth conformal future extension 
$\hat{M} \rightarrow M = \hat{M} \cup{\cal J}^+$, $\hat{g} \rightarrow g = \Omega^2\,\hat{g}$
and a diffeomorphism $j: S \rightarrow {\cal J}^+ = \{\Omega = 0\}$ that identifies $h$ with the metric
 induced 
by $g$ on ${\cal J}^+$  and $w$ with the ${\cal J}^+$-electric part of the extension of the rescaled 
conformal Weyl tensor $W^{\mu}\,_{\lambda \rho \nu}$ to $ {\cal J}^+$.}

\vspace{.1cm}

The result follows from an analysis of the constraints induced by the conformal field equations on a hypersurface $\{\Omega = 0\}$ and by solving a Cauchy problem backwards in time for the hyperbolic  reduced conformal fields equations. 
It is a semi-global result because the solutions are null-geodesically  future complete. 
All de Sitter-type vacuum solutions which admit smooth conformal extensions in the future (or past) are characterized here. 

No further restrictions on the topology of $S$, no smallness conditions on the data, and no restrictions on the conformal class of $(S, h)$ need to be imposed. 
The last property follows from the observation that the Hamiltonian constraint becomes trivial on the set $\{\Omega = 0\}$. This is related to the fact that
$(\lambda, S, \theta^2\,h_{ab}, \theta^{-1}\,w_{ab})$, with $\theta > 0$  a smooth function on $S$,
is also an asymptotic initial data set which determines the same `physical' solution as
$(\lambda, S, h_{ab}, w_{ab})$.
Because the solutions are obtained by solving Cauchy problems with data prescribed on 
${\cal J}^+ = \{\Omega = \,0\}$ the  {\it smooth conformal  extensibility of the solutions is built in here by  
construction}.

\vspace{.1cm}

 Let $\hat{S}$ denote a Cauchy hypersurface of a solution $(\hat{M}, \,\hat{g})$ as above and let  $(\lambda, \hat{S}, \hat{h}_{ab}, \hat{\chi}_{ab})$ be the standard  initial data induced by $\hat{g}$  on $\hat{S}$. We denote the set of such data  by ${\cal A}^+_S$. Thus
 ${\cal A}^+_S$ denotes the set of standard initial data sets which develop into de Sitter-type solutions that admit smooth conformal extensions in the future. 
 Similarly, ${\cal A}^{\pm}_S$ denotes the set of standard initial data sets which develop into de Sitter-type solutions that admit smooth conformal extensions in the past as well as in the future.

\vspace{.1cm}

\noindent
{\bf Strong future stability} \cite{friedrich:1986b}:
{\it The set ${\cal A}^+_S$ is open in ${\cal D}_S$ (in suitable Sobolev norm).}

\vspace{.1cm}

`Strong' has been added here to emphasize that not only geodesic (future) null completeness
but also smooth (future) conformal extensibility  is preserved under small perturbations.
The result is obtained  as follows. Let  $(\lambda, \hat{S}, \hat{h}_{ab}, \hat{\chi}_{ab})$ be an initial data set in  ${\cal A}^+_S$ and denote by  $(\hat{M}, \,\hat{g})$ the maximal, globally hyperbolic vacuum solution determined by it. Then $(\hat{M}, \,\hat{g})$ admits after a rescaling with a suitable conformal factor a smooth extension with boundary ${\cal J}^+$ in the future. Using the induced asymptotic data on 
${\cal J}^+$,  {\it it can be smoothly extended as a solution to the conformal field equations into a domain which contains a Cauchy hypersurface on which $\Omega = const. < 0$.}
 If $(\lambda, \hat{S}, \hat{h}^*_{ab}, \hat{\chi}^*_{ab})$ are data in ${\cal D}_S$ that are close (with respect to suitable Sobolev norms) to 
 $(\lambda, \hat{S}, \hat{h}_{ab}, \hat{\chi}_{ab})$ then data for the conformal field equations associated with these two data sets can be arranged so as to be also close to each other. The result then follows by Cauchy stability for the (symmetric) hyperbolic equations (see  \cite{kato:1975}) induced by the conformal field equations
 and the fact that equation (\ref{mat-B-Ric-scal-transf}) ensures that the sets $\{\Omega = 0\}$ are space-like hypersurfaces.

\vspace{.1cm}

\noindent
{\bf Strong global stability} \cite{friedrich:1986b}: {\it The set ${\cal A}^{\pm}_S$ is open in ${\cal D}_S$.} 

\vspace{.1cm}

This follows by similar arguments. It implies that the de Sitter solution or solutions obtain from it by factoring out suitable 
symmetries of $(\mathbb{S}^3, h_{\mathbb{S}^3})$
are  strongly  globally stable.
We note that in the two stability results  
 {\it the smooth conformal  extensibility of the solutions close to the reference solution is derived} as a consequence 
 of the conformal properties of Einstein's field equations. 

\vspace{.1cm}

\noindent
{\bf Generalizations including matter fields} \cite{friedrich:1991}:
{\it With suitably generalized definitions of
${\cal A}^{+}_S$, ${\cal A}^{\pm}_S$, ${\cal D}_S$  to include matter fields,
the results above generalize to Einstein's field equations (\ref{Einst-equ}) with $\lambda > 0$ coupled to 
`conformally well behaved' matter  field equations}.

\vspace{.1cm}

We shall not try to characterize `conformally well behaved' here because  it may require complicated transformations to arrive at a set of unknowns so that no $1/\Omega$ terms appear on  the right hand sides 
(\ref{mat-B-Ric-scal-transf}) to (\ref{mat-W-equ}) or in the transformed matter field equations and  the equations imply hyperbolic reduced systems. The statement certainly applies to matter equations  like the Maxwell and the Yang-Mills equations, discussed in detail in \cite{friedrich:1991}, which  in four space-time dimensions are conformally invariant  in the most direct sense. 
Other cases of matter fields with trace-free energy momentum tensor and conformally covariant equations have recently been worked out in 
\cite{luebbe:2013}, \cite{luebbe:valiente-kroon:2013}. Further below 
we shall discuss a less obvious case.

\vspace{.1cm}

\noindent
{\bf Generalizations to higher dimensions} \cite{anderson:2005a}: {\it The vacuum results above generalize to all even space-time dimension larger that four.}

\subsection{Questions, obstructions, numerical results}

The solutions which develop from data 
in ${\cal A}^+_S$ have been characterized indirectly  in terms of the asymptotic data induced  on ${\cal J}^+$ by the conformal extension of the solutions. A direct characterization
of the standard data in ${\cal A}^+_S$ is not known. That ${\cal A}^+_S$ is a proper subset of 
${\cal D}_S$ is illustrated by the analytically extended  Schwarzschild-de Sitter solutions, which
admit only patches of smooth conformal extensions in the future and past
\cite{gibbons:hawking:1977}. A more  extreme case is that of the standard Nariai space-time, given by

\hspace*{2.5cm}$\hat{M} = \mathbb{R} \times (\mathbb{S}^1 \times \mathbb{S}^2)$, 
$\quad \quad \hat{g} = - dt^2 + \cosh^2 t\,h_{\mathbb{S}^1} + h_{\mathbb{S}^2}$,

\noindent
which solves (\ref{vac-lambda}) with $\lambda = 1$. It  is globally hyperbolic and geodesically complete.
In  \cite{beyer:2009a}  has been used a topological argument
to show that {\it the standard Nariai solution does not even admit a patch of a smooth conformal boundary}. Observing that 
 $C_{\mu \nu \rho \lambda}[\hat{g}] \,C^{\mu \nu \rho \lambda} [\hat{g}]
= \Omega^4\,C_{\mu \nu \rho \lambda}|g]\,C^{\mu \nu \rho \lambda}|g]$
if $g_{\mu\nu} = \Omega^2\,\hat{g} _{\mu \nu}$,
where the contractions are performed with $\hat{g}$  on the left and with
 $g$ on the right hand side, and using the result
$ C_{\mu \nu \rho \lambda}[\hat{g}] \,C^{\mu \nu \rho \lambda}[\hat{g}]  = const. \neq 0$
of a calculation, 
it follows directly that there cannot exist a piece of ${\cal J}^{\pm}$ of class $C^2$.

Similarly, one would like to characterize the data in 
${\cal A}^+_S$ which are in fact  in ${\cal A}^{\pm}_S$. That the latter is a proper subset of 
${\cal A}^+_S$ is shown by the following examples. The space-time

\hspace*{2.5cm}$\hat{M} = \mathbb{R} \times \mathbb{T}^3$,  $\quad \quad \hat{g} = - dt^2 + e^{2\,t}\,k_0$,

\noindent
with $k_0$ an Euclidean metric on $\mathbb{T}^3$, solves (\ref{vac-lambda}) with $\lambda = 3$. It only admits a smooth extension in the future. Its  causal geodesics  are future complete but only the causal geodesics $t \rightarrow (t, p)$, $p \in \mathbb{T}^3$ are past complete. Another case is 
the space-time

\hspace*{2cm}$\hat{M} = \mathbb{R} \times (\mathbb{S}^1 \times \mathbb{S}^2)$, 
$\quad \hat{g} = - dt^2 + \sinh^2t\,h_{\mathbb{S}^1} + \cosh^2 t\,h_{\mathbb{S}^2}$

\noindent
which solves (\ref{vac-lambda}) with $\lambda = 3$,
admits smooth conformally extensions at the ends where $|t| \rightarrow \infty$ but becomes singular as $t \rightarrow 0$. These solutions illustrate a general phenomenon:

\vspace{.1cm}

\noindent
{\bf Obstructions  to smooth conformal extensibility in the past} \cite{andersson:galloway:2002}:
{\it A solution $(\hat{M}, \hat{g})$ to (\ref{vac-lambda}) with $\lambda > 0$ which develops  from data in 
${\cal A}_S^+$ does not even admit a patch of a smooth conformal extension in the past if the fundamental group of $S$ is not of finite order or if the asymptotic data $(h, w)$ induced by $ \hat{g}$ on ${\cal J}^+$ are such that 
the conformal structure defined by $({\cal J}^+, h)$ is not of positive Yamabe type.}

\vspace{.1cm}

The result gives little information about what exactly prevents  the solution from extending smoothly in the past. Further, there exist situations  in which smooth conformal extensibility fails for
reasons different from those given above. Let 
$\{\alpha^a\}_{a = 1, 2, 3}$ denote  a basis of 1-forms on $\mathbb{S}^3$ such that 
$\delta_{ab}\,\alpha^a\,\alpha^b = h_{\mathbb{S}^3}$ and 
$d\alpha^a =  - \epsilon_{bc}\,^a  \alpha^b   \wedge \alpha^c$.
Then 

\hspace*{.8cm}$\hat{M} = \mathbb{R} \times S^3$, 
$\quad \hat{g} = - \frac{1 + t^2}{v}\,d t^2 + (1+ t^2)(\alpha^1\,\alpha^1 + \alpha^2\,\alpha^2)
+ \frac{v}{1 + t^2}\,\alpha^3\,\alpha^3$,

\noindent
with $v = (1 + t^2)^2 - \alpha\,t$, $\alpha \in \mathbb{R}$,  denote  members of the  $\lambda$-Taub-NUT family which  solve (\ref{vac-lambda}) with $\lambda = 3$.
Their conformal structures extend smoothly in the past and in the future. With a suitable conformal scaling the metric $g$ induces on ${\cal J}^+ \sim \mathbb{S}^3$ 
the asymptotic initial data $h =  h_{\mathbb{S}^3}$,   
$w = - \frac{\alpha}{2}( \delta_{ab}\,\alpha^a\,\alpha^b - 3\,\alpha^3\,\alpha^3)$, so that the obstructions pointed out above are not present. 
If $|\alpha| <  \alpha_* \equiv  \sqrt{3}\cdot16/9$, then $v > 0$ for $\tau \in \mathbb{R}$, 
the solutions are geodesically complete and imply Cauchy data belonging to ${\cal A}^{\pm}_{\mathbb{S}^3}$.  If 
$|\alpha| \ge \alpha_*$ the solutions are no longer globally hyperbolic. 
If $\alpha =  \alpha_*$ 
the function $v$ has a double zero on the hypersurface
$\{ \tau = \tau_* \equiv 1/ \sqrt{3}\}$, which represents a smooth compact Cauchy horizon that contains closed null curve. If $\alpha >  \alpha_*$ the function $v$ has two simple zeros at values $\tau_{\pm}$ with  
$0 < \tau_- < \tau_* < \tau_+$. The hypersurfaces $\{\tau = \tau_{\pm}\}$ are Cauchy horizons 
which sandwich a domain that contains closed time-like curves.

For given asymptotic initial data  on $\mathbb{S}^3$ or $\mathbb{S}^1 \times \mathbb{S}^2$
Beyer investigates solutions to  the backward Cauchy problem for the conformal field equations 
numerically \cite{beyer:2007}. This allows him in particular to calculate solutions determined by data in ${\cal A}^{\pm}_{S}$ from ${\cal J}^+$ all the way down to ${\cal J}^-$. Among other solutions, he studies
a class of $\lambda$-Taub-NUT solutions larger than the one given above.
The investigation of the stability properties of solutions with Cauchy horizons (numerically a delicate adventure since  uniqueness of local extensions fails at Cauchy horizons) suggests that the solutions develop curvature singularities instead of Cauchy horizons if the asymptotic  data on ${\cal J}^+$
are slightly perturbed \cite{beyer:2008}.

\subsection{Extensions  beyond conformal boundaries}

The argument which give  the stability results uses the fact that solutions which admit smooth conformal extensions to ${\cal J}^+$ can in fact be smoothly extended as solutions to the conformal field equations into domains beyond ${\cal J}^+$ where $\Omega < 0$. 
The extension defines there another solution to the Einstein equations. 
The conformal representation of de Sitter space given by  $\Omega = \cos \tau $ and 
$g =  - d\tau^2 + h_{\,\mathbb{S}^{3}}$ extends analytically  to the manifold $\mathbb{R}\, \times \,\mathbb{S}^3$, where  $\Omega$ defines an infinite sequence of domains on which $\Omega \neq 0$,  separated by hypersurfaces where $\Omega = 0$, $d\Omega \neq 0$. 
On any such domain $\Omega^{-2}\,g$  is isometric to the de Sitter metric. Denote by ${\cal X} \sim \mathbb{S}^3$ a hypersurface which separates two of these domains and let $k > 0$ be an integer. Consider asymptotic vacuum data on ${\cal X}$.
Then, by Cauchy stability, the solution to the conformal field equations determined  by these data will extend over $k$ different domains where $\Omega \neq 0$ if the data are sufficiently close to the asymptotic de Sitter data induced  on  ${\cal X}$ by the metric $g$. In this case the `physical metrics' induced on the different domains will not necessarily be isometric.
It is an interesting question whether a solution to the conformal field equations which extends 
over an {\it infinite} number of domains where $\Omega \neq 0$ must necessarily be locally conformally flat.

So far the extensibility was used as a technical device. The `physical solutions' considered above, 
can hardly be considered as cosmological models, they expand exponentially in both time directions while 
current wisdom expects that the universe starts with a big bang.
One may still ask why the `physical world' should end at ${\cal J}^+$ 
if extensions across conformal boundaries are a natural consequence of the field equations.
At this stage we recall  that  matter fields with  non-zero rest mass have been ignored so far. There exists little precise information about the behaviour of the conformal structure near ${\cal J}^+_*$ if such fields are coupled to Einstein's equations. If the conformal structure can be controlled at all, it may depend in subtle ways on the specific nature of the matter model.

A transition process across conformal boundaries is at the basis of the 
 {\it conformal cyclic cosmology} proposed by Penrose \cite{penrose:2010}.
The underlying picture is that of a smooth, time oriented conformal structure of signature $(-,+,+,+)$ (which we shall refer to  as the {\it long conformal structure}) on a $4$-dimensional manifold ${\cal M} \sim \mathbb{R} \times S$ with  compact $3$-manifold $S$,  into which an infinite sequence of {\it aeons},  i.e. time oriented `physical'  solutions to Einstein's field equations with cosmological constant $\lambda > 0$, are conformally embedded so that any two consecutive aeons are separated by a {\it crossover $3$-surface} ${\cal X} \sim S$ which is space-like with respect to the conformal structure.
Each aeon starts with a big bang that `touches' the preceding crossover surface and ends in the future  with an  exponentially expanding phase for which the following 
crossover surface defines a smooth conformal boundary.

This scenario asks for a global PDE result which  
establishes the existence of a long conformal structure. The transition process through the crossover surfaces must be controlled with suitable versions of conformal field equations. 
 It must be clarified what happens at the prospective conformal boundaries in the presence of  fields with non-vanishing rest-masses. 
 In \cite{penrose:2010}  it is assumed that  only zero rest-mass fields will be present
in  some past neighbourhoods  of the crossover surfaces.  No justification is known
for this requirement. Finally, it  will not suffice to be able to glue  in specific cases  the future conformal boundary of a given aeon to
the hypersurface  that conformally represents the  isotropic singularity in the past of the subsequent aeon. 
A general mechanism is needed that forces the equations to take the route  from a expanding phase to big bang phase
instead of simply using  the smooth transitions across the conformal boundaries  discussed above.

Most likely, the last two problems are not independent of each other and it will be worthwhile to have a closer look at the asymptotic behaviour of solutions  involving non-zero-rest mass fields.
Ringstr\"om obtained quite general results on  the Einstein-scalar-field system.
Consider the de Sitter metric $\hat{g}$  together with a function $\phi = 0$ on 
$\hat{M} = \mathbb{R} \times \mathbb{S}^3$
as a special solution
to the Einstein's equations  (\ref{Einst-equ}) with energy momentum tensor 

\hspace*{2cm}$\hat{T}_{\mu \nu} = \hat{\nabla}_{\mu}\phi\,\hat{\nabla}_{\nu}\phi 
- \left(\frac{1}{2}(\hat{\nabla}_{\rho}\phi\,\hat{\nabla}^{\rho}\phi 
+ m^2\,\phi^2)+ U(\phi)\right)\,\hat{g}_{\mu \nu},
$

\noindent
coupled to the scalar field equation

\hspace*{4cm}$\hat{\nabla}_{\mu}\,\hat{\nabla}^{\mu} \phi = m^2\,\phi + U'(\phi)$,

\noindent
with rest-mass $m > 0$ and a smooth potential $U$. 

\vspace{.1cm}

\noindent
{\bf Global stability} \cite{ringstrom:2008}:
{\it If the  potential is such that  $U = O(|\phi|^3)$ as $\phi \rightarrow 0$,
then the  de Sitter solution (with $\phi \equiv 0$) is non-linearly stable in the sense  that 
sufficiently small perturbations of de Sitter Cauchy data  
on a hypersurface $\sim \mathbb{S}^3$
develop into  solutions to the coupled system whose 
causal geodesics are past and future complete and which are, with respect to suitable close norms, close  to the de Sitter solution.}

\vspace{.1cm}

The work referred to above gives detailed estimates in terms of the physical metric.
Getting precise ideas about the asymptotic behaviour of the conformal structure will require more, however. 
In studying these questions it may be useful to take into account the possibility that 
a loss of differentiability  of the long conformal structure  at the crossover surfaces
 may be admissible as long as uniqueness of the extensions is assured. 
 The loss may depend on various specific features of the matter model.
The following, somewhat unexpected, result shows that there are possibilities which are not obvious if one just  looks at the Einstein equations in their standard form. 

\vspace{.1cm}

\noindent
{\bf Strong global stability} \cite{friedrich:non-zero rest-mass}:
{\it If the mass and the cosmological constant  are related   in the Einstein-scalar-field system by 
$3\,m^2 = 2\,\lambda$ and the potential satisfies  $U = O(|\phi|^4)$ as 
$\phi \rightarrow 0$, then  Cauchy data sufficiently close to de Sitter data with $\phi = 0$ evolve into globally hyperbolic solutions that admit smooth conformal extensions in the past and in the future. }

\vspace{.1cm}

If the scalar field is replaced by the function $\psi = \Omega^{-1}\,\phi$, the conformal field equations and the transformed scalar field equation contain under the assumptions above no  $1/\Omega$ terms 
and imply in fact hyperbolic reduced systems. With asymptotic data on a slice 
$\{\Omega = 0\}$ that generalize the asymptotic vacuum data considered earlier, a local existence result follows and arguments similar to the ones given earlier imply the result.

This raises various questions. To which extent can the smoothness assumptions and results be relaxed and the 
matter models be generalized ?  Can extensions of low smoothness and suitable matter models help  initiate the transition process from expanding to big bang phases ? Do there exist physical fields satisfying the condition ? Does the relation between the matter field and the  cosmological constant 
shed any light on the origin and the role of the cosmological constant ? These questions will have to be discussed elsewhere.

\section{Minkowski-type solutions.}
\label{vanishing lambda}

The calculation of the gravitational radiation generated by spatially  localized processes  like the encounters of stars or merges of black holes is one of the main motivations for analyzing  asymptotically flat solutions.
The simply connected, conformally flat model case is Minkowski-space, given in spatial spherical coordinates by

\vspace{.1cm}

\hspace*{2cm}$\hat{M} = \mathbb{R}^4$, $\quad \quad \hat{g} = -dt^2 + dr^2 + r^2\,h_{\mathbb{S}^2}$.

\vspace{.1cm}

\noindent
Coordinates $\tau$ and $\chi$ and a conformal factor $\Omega$ satisfying 

\vspace{.1cm}

\hspace*{.5cm}$t = \Omega^{-1}\, \sin \tau, \quad r = \Omega^{-1}\,\sin \chi, \quad  
\Omega  = \cos \tau + \cos \chi, \quad 0 \le \chi, \quad |\tau \pm \chi| < \pi$,

\vspace{.1cm}

\noindent
give $\Omega^2\,\hat{g} = h \equiv -d\tau^2 + d\chi^2 + \sin^2 \chi\,h_{\mathbb{S}^2}$, whence a smooth conformal  embedding of Minkowski space into the Einstein cosmos
$(M^* = \mathbb{R} \times \mathbb{S}^3, g^* = -d\tau^2 + h_{\mathbb{S}^3})$ \cite{penrose:1965}.

The metric $g$ and the conformal factor $\Omega$ extend smoothly to the
range $0 \le \chi$,  $|\tau \pm \chi| \le \pi$  of the coordinates. This extension adds to $\hat{M}$ the sets
${\cal J}^{\pm} = \{\tau = \pm\,(\pi - \chi), \,\,0 < \chi < \pi\} $, which are null hypersurfaces for the extended metric $g$ referred to as future and past null infinity. 
The 
sets  $i^0 = \{\tau = 0, \chi =\pi\}$  and  $i^{\pm} = \{\tau = \pm \pi, \chi =0\}$
represent regular points of  the conformal extension, it holds there $\Omega = 0$, $d\Omega = 0$, and $0 \neq Hess_g\Omega \sim g$.
For a given space-like slice  $\hat{S}_{t_o} = \{\sin \tau = t_o\,\Omega\}$, $t_o \in \mathbb{R}$, one may consider  $i^0$ as a point added at spatial infinity 
which makes the slice  into  a sphere $\sim \mathbb{S}^3$. 
The point  $i^0$ defines, however an endpoint (in both directions)
for all space-like geodesics and thus represents space-like infinity for 
$(\hat{M}, \hat{g})$. The points $i^{\pm}$ are approached by the time-like geodesics in the future and the past and thus represent for $(\hat{M}, \hat{g})$ future and past time-like infinity.
The set ${\cal J}^+$ is ruled by the past directed null geodesics through the point $i ^+$, which coincide with future directed null geodesics through $i^0$. Similar relations hold for 
${\cal J}^-$, $i^-$, and $i^0$.

There are of course no stars or black holes around here but since the registration of gravitational radiation takes place at large distances from the sources, 
generalizations of the situation above to the far fields of non-trivial vacuum solutions are of particular interest. It is sometimes said that it were too extreme an idealization to put the measuring device at null infinity or, in other words, to read off the radiation field there.
This is in fact one of the questions we are interested in when we try to control the field near null infinity.
If the field turns out to admit a smooth conformal extension at null infinity then the size and structure of the field will hardly be affected if the location of the ideal measuring device at null infinity is shifted slightly,  in terms of the $g$ adapted conformal coordinates, into the space-time. In terms of the physical metric $\tilde{g}$ such a shift covers an infinite distance, which puts the measuring device at a reasonable distance to the sources.  This is the situation considered in most numerical calculations.
What is lost in this procedure, however,  is the unique tangent space of null infinity which serves to define the radiation field and gives an automatable prescription for the numerical calculation of radiation fields.

Quite early  the requirements of Definition \ref{conf-ext} were shown to be met by some of the 
most important explicit vacuum solutions that admit time-like Killing fields \cite{hawking:ellis:1973},
\cite{penrose:1965}. These were special cases of a general fact.

\vspace{.1cm}

\noindent
 {\bf Static and stationary fields}  \cite{dain:stationary}: {\it Asymptotically flat static or stationary vacuum solutions admit smooth conformal extensions at future and past null infinity}.

\vspace{.1cm} 
 
This result establishes the existence of a fairly large class of asymptotically simple vacuum solutions.
The main purpose of introducing  Definition \ref{conf-ext} is, however, to discuss gravitational  radiation  and all the solutions above have vanishing radiation fields. There remains the question to what extent Definition \ref{conf-ext}  applies to dynamical solutions.

In hindsight the PDE problem whose analysis paved the way to the notion of asymptotic simplicity can understood  as the {\it asymptotic characteristic initial value problem} where data are prescribed on an outgoing null hypersurface ${\cal N}$ which is supposed to intersect future null infinity in a 2-dimensional space-like slice $\Sigma$ and the part ${{\cal J}^+}'$ of future null infinity in the past of $\Sigma$.
A detailed formulation of this problem for the conformal field equations  which specifies the freedom to prescribe data has been given in \cite{friedrich:asymp-sym-hyp:1981b}. Its elaboration gives:

\vspace{.1cm}

\noindent
{\bf  Well-posedness of the  asymptotic characteristic initial value problem}  \cite{kannar:1996a}: {\it For given smooth {\it null data} on ${\cal N}$ and ${{\cal J}^+}'$ and certain smooth functions given on  $\Sigma$, there exists a smooth solution to the conformal vacuum field equations 
in a past neighborhood ${\cal U}$ of $\Sigma$ which induces the given data on ${\cal U} \cap ({\cal N} \cup {{\cal J}^+}')$. It induces on $\hat{{\cal U} } = {\cal U} \setminus  {{\cal J}^+}'$ a unique solution
$\hat{g}$ to Einstein's vacuum field equation $\hat{R}_{\mu\nu} = 0$}.

\vspace{.1cm}

The smooth conformal extensibility of $(\hat{{\cal U} }, \hat{g})$  has been built in by the way the PDE problem is formulated.
The freedom to prescribe null data (two components of $W^{\mu}\,_{\nu \rho \lambda}$ on each hypersurface) is similar to that in characteristic initial value problems for Einstein's vacuum field equations with data given on two intersecting null hypersurfaces 
${\cal N}_1$, ${\cal N}_2$ which are thought as lying  in the physical space-time \cite{friedrich:1981a}. 
 Only some differences in the freedom to prescribe data on 
 $\Sigma$ resp. ${\cal N}_1 \cap  {\cal N}_2$ indicates that ${{\cal J}^+}'$ is geometrically a special hypersurface. The null datum on ${{\cal J}^+}'$ is in fact the radiation field $\psi_0$.
  If the data on ${\cal N}$ and $\Sigma$ are trivial, the solution is completely determined by $\psi_0$. In the time reversed situation it is thus quite natural to identify $\psi_0$ with the incoming radiation field on past null infinity. Since the radiation field can be prescribed freely we have solutions of the type we are looking for.
 There also exists a {\it real analytic version} of this result \cite{friedrich:1982}. The solution to the conformal field equations then extends analytically  into the future of null infinity to a domain where $\Omega < 0$.

The characteristic initial value problem for the conformal field equations whose solutions supplies  
{\it purely radiative space-times} is of particular interest \cite{friedrich:pure rad}.  These space-times are defined
be the requirement that they posses a smooth past conformal boundary ${\cal J}^-$ whose null generators are complete in a conformal gauge that makes ${\cal J}^-$ expansion free and which admit a smooth conformal extension  containing a point  $i^-$ so that 
 ${\cal N}_{{i^-}} = {\cal J}^- \cup \{i^-\}$ is the cone generated by the future directed null geodesics emerging from  $i^-$. Some authors say  `purely radiative' under much weaker assumptions. It should be noted, however,  that only with sufficient regularity at $i^-$ the solutions will be determined uniquely by the radiation field, a  counter example being given by the Schwarzschild solution.
 The  simplest way to create purely radiative space-times is to assume that the radiation field vanishes in a neighbourhood  of $i^-$, so that the solution will be Minkowskian near past time-like infinity.
If one wants to exploit the full freedom to prescribe data, however, one has  to face problems arising from the non-smoothness of the initial set at the vertex. One needs an appropriate notion of smoothness for the free data on ${\cal N}_{{i^-}}$ and it must be shown that  field equations themselves then ensure the smoothness of the solution in the future of  ${\cal N}_{{i^-}}$ \cite{friedrich:taylor-at-i}. The most difficult part, the existence problem near $i^-$, has been solved 
only recently.

\vspace{.1cm}

\noindent
{\bf Existence for the pure radiation problem near  past time-like infinity $i^-$} \cite{chrusciel-paetz-2013}:
{\it For a given radiation field on the cone ${\cal N}_{{i^-}}$ that satisfies appropriate smoothness conditions, there exists near $i^-$ a unique (up to diffemorphisms) smooth solution to the vacuum equations  in the future of 
${\cal N}_{{i^-}}$. For this solution ${\cal N}_{{i^-}}$ represents a smooth conformal past boundary with regular vertex $i^-$ on which the solution induces the given data.}

\vspace{.1cm}

Characteristic problems are important in various arguments  and are being used as the basis 
of numerical calculations extending to null infinity. They will not be considered here any further. Being  ruled by null geodesics, null hypersurfaces have an intrinsic tendency to develop caustics that can cause extreme technical difficulties.

\subsection{The hyperboloidal initial value problem.}

We shall consider  initial value problems based on space-like hypersurfaces. Two classes 
of such hypersurfaces are of interest to us.
These are {\it standard Cauchy hypersurfaces} like the sets $\hat{S}_{t_o}$  considered above, that extend to space-like infinity, and  {\it hyperboloidal hypersurfaces}  like the sets 
$\{\tau = \tau_o, 0 \le \chi < \pi - |\tau_o|\}$, $0 < |\tau_o| =  const. < \pi$, 
 in the conformally extended Minkowski space, 
which {\it extend smoothly to null infinity  as space-like slices}. The prototype example, obtained for 
$\tau_o = \frac{\pi}{2}$, is the unit hyperboloid  $ H^+ = \{ - t^2 + r^2 = -1, t > 0 \}$ which motivates  the name  \cite{friedrich:1983}. 
In the conformally extended  Minkowski space it is easily seen that the future null cone 
with vertex at the origin, future null infinity ${\cal J}^+$,  and the extension of the hypersurface $ H^+ $ 
 intersect each other transversally.  Referring to  the latter as
`asymptotically null', as it is done sometimes,  is thus easily misleading.

Hyperboloidal initial data $( \hat{S}, \hat{h}_{ab}, \hat{\chi}_{ab})$ for Einstein's vacuum field equations 
$\hat{R}_{\mu\nu} = 0$ 
have in common with asymptotically flat Cauchy data that 
the underlying Riemannian space  $( \hat{S}, \hat{h}_{ab})$ is required to be orientable and geodesically complete, the vacuum constraints
$R[\hat{h}] = \hat{\chi}_{ab}\,\hat{\chi}^{ab} - (\hat{\chi}_a\,^a)^2$ and 
$\hat{D}_b\,\hat{\chi}_a\,^b = \hat{D}_a\,\hat{\chi}_b\,^b$ must be satisfied, and 
the mean extrinsic curvature $\hat{\kappa} = \hat{\chi}_a\,^a$ can be assumed to be constant.  
This will be done in the following to simplify the discussion.
They differ, however,  in their asymptotic behaviour.
In the hyperboloidal case one must then have $\hat{\kappa} \neq 0$
whereas $\hat{\kappa} = 0$ in the asymptotically flat case. Moreover, if hyperboloidal data are supposed to lead to a asymptotically smooth situation, they must admit a smooth conformal completion 

\vspace{.1cm}

\hspace*{1cm}$\hat{S} \rightarrow S = \hat{S} \cup \Sigma, \quad
\hat{h}_{ab} \rightarrow h_{ab} = \Omega^2\,\hat{h}_{ab}, \quad
\hat{\chi}_{ab} \rightarrow \chi_{ab} = \Omega\,(\hat{\chi}_{ab} - \frac{1}{3}\,\hat{\kappa}\,\hat{h}_{ab})$,

\vspace{.1cm}

\noindent
where
 $(S, h_{ab})$ is a Riemannian space with compact boundary $\partial S = \Sigma$, 
$\Omega$ a smooth defining function of $\Sigma$ with $\Omega > 0$ on $ \hat{S}$,
and  
$W^{\mu}\,_{\nu \lambda \rho} = \Omega^{-1}\, \hat{C}^{\mu}\,_{\nu \lambda \rho}$ 
extends smoothly to $\Sigma$ on $S$ where $ \hat{C}^{\mu}\,_{\nu \lambda \rho}$ denotes then conformal Weyl tensor determined by the metric $ \hat{h}_{ab}$ and the second fundamental form $ \hat{\chi}_{ab}$. {\it Such data will be called smooth}.

\vspace{.1cm}

\noindent
{\bf Existence for hyperboloidal problems}  \cite{friedrich:1983}: {\it Smooth hyperboloidal initial data develop into a unique smooth, maximal, globally hyperbolic  solution to Einstein's vacuum field equations $\hat{R}_{\mu\nu} = 0$ which admits in the future of the embedded hypersurface $\hat{S}$ a smooth conformal extension at future null infinity, that approaches  $\Sigma$ in its past end,  and which possesses a Cauchy horizon in the past of $\hat{S}$  that approaches $\Sigma$ in its future end.}

\vspace{.1cm}

While in the solution above the future directed null geodesics  which approach future null infinity are future complete,  the smooth  hyperboloidal hypersurfaces that connect the two future null infinities  of the conformally extended Schwarzschild-Kruskal space-time \cite{hawking:ellis:1973} show that further assumptions are required to ensure future completeness for all null geodesics.

Hyperboloidal initial data for which this completeness requirement is satisfied are
given by the data induced on a space-like hypersurface $\hat{S}$  in Minkowski space 
with the following properties: Any past inextendible causal curve starting in the future of $\hat{S}$ intersects $\hat{S}$ precisely once and $\hat{S}$ has a (unique) smooth space-like extension $S$ in the conformally extended Minkowski space so that the surface
$\Sigma = S \cap {\cal J}^+$ is  diffeomorphic to $\mathbb{S}^2$.
Data of this type will be referred to as {\it Minkowskian hyperboloidal data} and their future development 
as a {\it Minkowskian hyperboloidal development}.

\vspace{.1cm}

\noindent
{\bf Strong future stability for Minkowskian hyperboloidal developments}  \cite{friedrich:1986b}:
{\it Let $( \hat{S}, \hat{h}^*_{ab}, \hat{\chi}^*_{ab})$ denote Minkowskian hyperboloidal data 
with smooth conformal extension to the $3$-manifold  $S = \hat{S} \cup \Sigma$.
Then any smooth hyperboloidal vacuum initial data set $( \hat{S}, \hat{h}_{ab}, \hat{\chi}_{ab})$  which is sufficiently close (in suitable Sobolev norms) to the Minkowskian data set develops into a solution which is null geodesically future complete and admits a smooth conformal extension at future null infinity with conformal boundary ${\cal J}^{+'}$. Moreover, the conformal extension contains a regular point $i^+$ 
such that the past directed null geodesics emanating from  $i^+$ generate the set
${\cal J}^{+'}$ and approach the boundary $\Sigma$ attached to $\hat{S}$ in their past.}

\vspace{.1cm}

The proof uses again the possibility to extend solutions to the conformal field equations through sets where $\Omega$ vanishes. We don't go into any details but point out the remarkable consequence of equations  (\ref{mat-B-Ric-ten-transf}) and  (\ref{mat-s-equ}) that
 the null generators of the set ${\cal J}^{+'}$ are forced to meet under the given conditions at exactly one point $i^+$, where the conformal factor has a non degenerate singularity with $Hess_g\Omega = s\,g$, 
$s = s|_{i^+}  \neq 0$.

\vspace{.1cm}

\noindent
{\bf Generalizations including matter fields} \cite{friedrich:1991}: {\it The vacuum results above can be generalized to include conformally well behaved matter fields coupled to Einstein's equations.}

\vspace{.1cm}

\noindent
{\bf Generalization to higher dimensions} \cite{anderson:chrusciel:2005}: {\it The vacuum results generalize to even space-time dimensions larger than four}.

\vspace{.1cm}

Hyperboloidal initial data can be constructed by a suitable adaption  of the conformal method known from the construction of asymptotically flat Cauchy data. Certain {\it seed data consistent with the required fall-off behaviour at space-like infinity} are prescribed freely and then elliptic equations are solved to determine correction terms so that the corrected data will satisfy the constraints.

\vspace{.1cm}

\noindent
{\bf Existence of smooth hyperboloidal vacuum data}
\cite{andersson:chrusciel:1996}, \cite{andersson:chrusciel:friedrich:1992}:
{\it Seed data on a $3$-manifold $S$ with boundary $\Sigma$ which extend smoothly to  $\Sigma$
determine solutions to the vacuum constraints. In general these
admit at $\Sigma$ only asymptotic expansions in terms of powers of $x$ and $\log x$ where $x$ is a local coordinate with $x = 0$ on $\Sigma$  and $x > 0$ on  $\tilde{S} = S \setminus \Sigma$ (polyhomogenous expansions). The hyperboloidal data are smooth if and only if the seed data satisfy certain conditions at  $\Sigma$.}

\vspace{.1cm}

It is important here to note  that conditions on the seed data need only  be imposed  at the boundary at infinity  to ensure the smoothness of the resulting hyperboloidal data. 
The result suggests a generalization.

 \vspace{.1cm}

\noindent
{\bf Existence of polyhomogeneous hyperboloidal vacuum data} \cite{andersson:chrusciel:1996}:
{\it Seed data which are smooth on $\hat{S}$ and admit certain polyhomogenous expansions at  
$\Sigma$
determine hyperboloidal vacuum data which admit polyhomogenous expansions at  $\Sigma$.}

\vspace{.1cm}

The fact that the data are `physical'  only on the open set $\hat{S}$  leaves {\it  a large ambiguity of how to specify the data at $\Sigma$ if some roughness is admitted}. 
It is a critical  open question, which will come up again in the standard Cauchy problem, whether there exist physical situations of interest which can only be modeled by using non-smooth data. 

\vspace{.1cm}

For the calculation of radiation fields at null infinity the hyperboloidal initial value problem is as good as the standard Cauchy problem and it has the advantage of avoiding the difficulties at space-like infinity discussed  below. P.  H\"ubner pioneered
the numerical calculation  of solutions including parts of future null infinity   from hyperbolical data 
 \cite{huebner:1996}. He developed a code for  the conformal vacuum field equations which enabled him to calculate numerically,  without assuming any symmetries, the entire future of the hyperboloidal initial slice  as considered in the strong stability result above. The calculation 
  includes the radiation field on the asymptotic region ${\cal J}^{+'}$ and its limit at $i^+$  
 \cite{huebner:2001}. 
A survey on this and other numerical  developments involving the conformal field equations is given by Frauendiener \cite{frauendiener:2004}.
Once it had been established that certain classes  of hyperboloidal data develop into solutions with a smooth asymptotic structure it became less daring to try other equations.
There are now being developed  numerical codes based on equations that are more directly related to the singular conformal representation (\ref{Ric[g]})  \cite{rinne:moncrief:2013}.

\subsection{The standard Cauchy problem.}

For asymptotically flat space-times the standard Cauchy problem for Einstein's field equations  is more fundamental than the hyperboloidal problem, its solutions cover the past as well as the future in a unified way while a solution to a hyperboloidal problem is thought of  as  part of an ambient asymptotically flat space-time.
One may wonder whether  the problem raised above about non-smooth hyperboloidal initial data could be answered by analyzing the developments of standard Cauchy data. The picture of the conformally compactified Minkowski space and the results on the hyperboloidal initial value problem then suggest  that the  field equations decide on the asymptotic  smoothness at null infinity in any neighbourhood of space-like infinity.  There is again an ambiguity concerning the fall-off behaviour.

One has to decide between conflicting requirements. Too much generality can obscure important features by irrelevant noise while overly stringent conditions aiming at  sharp control  of  physical concepts may cause a loss of physically relevant input. To make a useful choice  one needs to understand a reasonably broad spectrum of possibilites.
 
 Existence results are usually formulated in terms of suitable function spaces which encode in a precise way  the fall-off behaviour and other properties of the solutions. In the following comparisons we shall ignore all these important technicalities and just use order symbols to indicate the asymptotic behaviour of the fields which is essential to our discussion. For the full precision we refer to the original articles. 
In the following we consider smooth initial data sets, i.e. solutions  
$(\hat{S},  \hat{h}_{ab},  \hat{\chi}_{ab})$  to the vacuum constraints, on the manifold
$\hat{S} =  \mathbb{R}^3$ with standard Euclidean coordinates $\hat{x}^a$,
which are in a suitable sense close to the  Minkowski data  $(\hat{S},  \delta_{ab},  0)$. 
Bieri obtained a quite general  result.

\vspace{.1cm}

\noindent
{\bf Stability of Minkowski space A} \cite{bieri:zipser:2009}: {\it Smooth vacuum Cauchy data 
$(\hat{S},  \hat{h}_{ab},  \hat{\chi}_{ab})$ which are sufficiently close to the Minkowski data
and which are asymptotically flat so that  

\vspace{.1cm}

\hspace{2cm}$\hat{h}_{ab} = \delta_{ab} + o_3(|\hat{x}|^{-1/2}), \quad \quad 
 \hat{\chi}_{ab} = o_2(|\hat{x}|^{-3/2})$,

\vspace{.1cm}

\noindent
develop into solutions to Einstein's vacuum equations $\hat{R}_{\mu \nu} = 0$, for which all causal geodesics are complete and whose curvature tensor approaches zero  asymptotically in all 
directions.}

\vspace{.1cm}

A global stability result based on such weak fall-off conditions is a remarkable mathematical achievement. One has to pay a price, however. The analysis gives little 
information on the precise behaviour of the solution near null infinity. In particular, {\it a concept of radiation field is no longer available} \cite{bieri:2009}.
Christodoulou and  Klainerman obtained their result under  stronger requirements.

\vspace{.1cm}

\noindent
{\bf Stability of Minkowski space B} \cite{christodoulou:klainerman:1993}:
{\it Smooth vacuum Cauchy data 
$(\hat{S},  \hat{h}_{ab},  \hat{\chi}_{ab})$ which are sufficiently close to the Minkowski data
and which are asymptotically flat so that  

\vspace{.1cm}

\hspace*{1.7cm}$\hat{h}_{ab} = \left(1 + 2\,m\,|\hat{x}|^{-1}\right)\,\delta_{ab} + o_4(|\hat{x}|^{-3/2}), \quad \quad  \hat{\chi}_{ab} = o_3(|\hat{x}|^{-5/2})$,

\vspace{.1cm}

\noindent
with some constant $m > 0$ develop into solutions to Einstein's vacuum equations $\hat{R}_{\mu \nu} = 0$, for which all causal geodesics are complete and whose curvature tensor approaches zero  asymptotically in all 
directions. Null infinity is complete. A concept of radiation field can be defined}.

\vspace{.1cm}

Some information about null infinity, such as being  approximated by complete null geodesics, is obtained,  but sharp fall-off  statements are missing. The  results obtained on the fall-off behaviour of the conformal Weyl tensor at null infinity are weaker than those  required by the Sachs peeling behaviour. If they are sharp the solutions do not admit smooth conformal extensions.
This raises doubts  whether there exist asymptotically flat vacuum data at all that develop into solutions which are null geodesically complete and admit  smooth conformal extensions at null infinity \cite{christodoulou:klainerman:1993}.  The characterization  of such data clearly requires a detailed analysis  of the evolution near space-like infinity. 
It is reasonable to base the investigation on a choice of data with clean fall-off behaviour at all orders.

\vspace{.1cm}

\noindent
{\bf Existence of Cauchy data with prescribed asymptotic behaviour }  \cite{dain:friedrich}:
{\it There exists a large class of asymptotically flat vacuum Cauchy data on $\mathbb{R}^3$ so that in suitable coordinates $\hat{x}^a$ near space-like infinity the data behave as

\vspace{.1cm}

\hspace*{1cm}$\hat{h}_{ab} = \left(1 + 2\,m\,|\hat{x}|^{-1}\right)\,\delta_{ab} + O(|\hat{x}|^{-2}), \quad \hat{\chi}_{ab} = O(|\hat{x}|^{-2}) \quad \mbox{as} \quad |\hat{x}| \rightarrow \infty$,

\vspace{.1cm}

\noindent
and admit  asymptotic expansions in terms of powers of 
$|\hat{x}|^{-1}$ as $|\hat{x}| \rightarrow \infty$  with smooth, bounded coefficients.}

\vspace{.1cm}

In  \cite{dain:friedrich} are prescribed seed data  possessing this property, where  
the seed metric is required in addition to admit a smooth conformal compactification at space-like 
infinity (for possible generalizations see  \cite{dain:stationary}).
It is then shown that the elliptic equations give correction terms with the desired asymptotic expansion
if and only if $|\hat{x}|^3\,\hat{\chi}_{ab}$ is bounded on $\mathbb{R}^3$. 
Otherwise one obtains data which only admit polyhomogeneous expansions. It is interesting to note that {\it the latter are admitted  in} \cite{bieri:zipser:2009}
{\it but excluded in} \cite{christodoulou:klainerman:1993}. 
{\it They will also be excluded in the following discussion}.

The smooth conformal extension of Minkowski space contains  the point $i^0$ that represents space-like infinity for the space-time as well as for any Cauchy data.
If one stipulates a similar picture  for general asymptotically flat vacuum solutions 
with mass $m > 0$, one finds that the data for the conformal field equations induced on Cauchy hypersurfaces are strongly  singular at the point $i^0$. In the following it will become clear that a detailed and general analysis of the field equations and their evolution properties  is  impossible if the structure near space-like infinity  exhibited below  is compressed into one point. 

Surprisingly, the Einstein equations allow us to define a setting which explains why
a stability result as strong and unrestricted as in the de Sitter case cannot be obtained in the 
case $\lambda = 0$. 

\vspace{.1cm}

{\bf The regular finite Cauchy problem} \cite{friedrich:i-null}:
{\it With asymptotically flat initial data as above the Cauchy problem for $\hat{R}_{\mu\nu} = 0$ is equivalent to an initial value problem for the  general conformal field equations with smooth initial data  on a 3-manifold  $S$ with boundary  $I^0 \sim \mathbb{S}^2$. These data develop smoothly on a manifold $M$ diffeomorphic to an open neighborhood of $S \equiv \{0\} \times S$ in $\mathbb{R} \times S$ which, assuming  a suitable `time'-coordinate $t$ taking values in the first factor, has $I =\, ]-1, 1[ \times I^0$ as a boundary.}

\vspace{.1cm}

It is a consequence of the assumptions  that the data extend in a suitable conformal scaling smoothly to the boundary $I^0$, which represents space-like infinity on the initial slice $S$.
What may look strange, is that the Cauchy problem has been replaced  by an initial-boundary value problem. The boundary is, however,  of a very peculiar nature. It is `totally characteristic' in the sense that at points of the boundary the hyperbolic reduced equations contain only differential operators tangential to the boundary. 
On the boundary the unknowns are thus  evolved by  inner equations and boundary data  cannot be prescribed. 
The set $I$, which is generated in the given gauge by the evolution process  from $I^0$,
 defines a smooth extension of the physical manifold which
represents space-like infinity for the solution space-time.

The setting has further remarkable features. In the given gauge the conformal factor $\Omega$ is an explicitly known function of the coordinates and there are sets 
 ${\cal J}^{\pm}_{\#} = \{\Omega = 0, d\Omega \neq 0\}_{\pm}$ with known finite (gauge dependent) coordinate location  in the future  and the past of $S$ respectively so that $\Omega > 0$ between these sets on $M \setminus I$.
If the solution to the conformal field equations extends to these sets with sufficient smoothness they will in fact define the conformal boundary at null infinity so that ${\cal J}^{\pm}_{\#} = {\cal J}^{\pm}$.

The sets   ${\cal J}^{\pm}_{\#} $  approach the boundary $I$ transversely at the {\it critical sets} $I^{\pm} = \{\pm 1\} \times I^0$, which can be understood as defining a boundary of $I$. While the reduced equations will be hyperbolic along $I$ as well as  between 
 ${\cal J}^{-}_{\#} $ and  ${\cal J}^{+}_{\#} $  (as long as the gauge does not break down), and even on the latter sets if the solution extends smoothly, their hyperbolicity degenerates in a specific way at the critical sets.  {\it The decision which initial data evolve into solutions that admit a smooth  conformal structure at null infinity takes place precisely at these sets}. 

Data which are  static in a neighborhood  of $I^0$ in $S$ evolve into solutions  which are real analytic and   extend with that property across $I^{\pm}$ and  ${\cal J}^{\pm}$ \cite{friedrich:cargese}. If the data are stationary the setting works  as well and one gets a similar result \cite{acena:kroon}.
For more general data the loss of hyperbolicity at $I^{\pm}$ can entail, however,  a loss of smoothness at these sets. 

\vspace{.1cm}

\noindent
{\it This is a fundamental difference with the hyperboloidal initial value problem. While in the latter  smoothness of the initial data ensures smoothness of the conformal boundary near the initial slice, this is not the case in the standard Cauchy problem. Here, the field equations define a hierarchy of conditions,  involving all orders of differentiabilty,  which go beyond simple smoothness requirements on the initial slice.}

\vspace{.1cm}

\noindent
Denote by $u$ the set of unknowns in the conformal field equations, by $r$ a local coordinate near $I$ with $r = 0$  on $I$ and $r > 0$ on $M \setminus I$, and by $t$ a `time' coordinate on $I$ with $t = 0$ on $I^0$ and $t = \pm 1$ on $I^{\pm}$. Then the peculiar nature of the boundary $I$ has the consequence that the functions $\partial_r^pu$, $ p \ge 1$,  evolve from the initial data $\partial_r^pu|_{I^0}$ as solutions to {\it linear, hyperbolic transport equations in $I$}. Explicit calculations show that these functions, which are smooth  on $I$, do in general  not extend smoothly to $I^+$ but only admit asymptotic expansion in terms of
$(t - 1)^k \, \log (t - 1)^l$ as $t \rightarrow 1$ with coefficients that are smooth functions on $I^0$ and with exponents $k$ which are increasing with $p$ so that the singular behaviour  gets milder at higher order. A similar behaviour is found at $I^-$.

{\it In which way does this affect the smoothness at the sets ${\cal J}^{\pm}_*$ ?} This question has not been answered yet. But that it will have an effect is seen if the setting is linearized at Minkowski space. The resulting problem is then controlled by the Bianchi equation
 for the linearized rescaled conformal Weyl tensor $W'^{\mu}\,_{\nu \rho \lambda}$. It turns out that if the initial data are such that some quantity $\partial_r^p W'^{\mu}\,_{\nu \rho \lambda}$ on $I$ 
 develops a logarithmic term at $I^+$ then the logarithmic singularity spreads along the null generators of null infinity \cite{friedrich:spin-2}. The situation can hardly be expected to improve in the non-linear case.

{\it Which conditions on the initial data on $S$ ensure that the functions 
$\partial_r^pu$ extend smoothly to $I^{\pm}$ } ? This question is technically quite difficult.
A  first family  of such conditions have been derived in  
\cite{friedrich:i-null}, which concentrates on the time reflection symmetric case so that the physical data on $S \setminus I^0$ are given by the $3$-metric $\hat{h}_{ab}$. Let $B_{ab}$ denote he (dualized) Cotton tensor of the rescaled $3$-metric $h_{ab}$. It has been shown in \cite{friedrich:i-null}
that  the symmetrized $h$-covariant derivatives  of $B_{ab}$ {\it necessarily} vanish at all orders at space-like infinity if  the $\partial_r^pu$ extend smoothly to $I^{\pm}$. 
Since the sequence of these conditions is conformally invariant they define a condition on the asymptotic conformal structure of $\hat{h}_{ab}$. If the metric $\hat{h}_{ab}$ is conformally flat near space-like infinity these conditions are satisfied. In that case the metric is determined near space-like infinity  only the by scaling factor of the flat metric, which is restricted by the Hamiltonian constraint.
It has been shown that even in that case there might arise obstructions to the smoothness of the 
$\partial_r^pu$ at $I^{\pm}$ \cite{j.a.v.kroon:2004a}.

In the time  reflection symmetric case the complete set of necessary conditions is not known yet 
and the situation is even less clear in general.
In the time reflection symmetric case there are, however, strong  indications which suggest that 
a {\it necessary and sufficient condition 
for the smoothness of the $\partial_r^pu$, $ p \in \mathbb{N}$,  is that the datum  
$\hat{h}_{ab}$ is asymptotically static at space-like infinity}. We refer to \cite{friedrich:conf-str-static-data} for a detailed discussion.

\subsubsection{Asymptotically special initial data}

The discussion above raises questions about the possibilities to construct solutions to the vacuum constraints which satisfy at space-like infinity additional asymptotic conditions at all  orders. Cutler and Wald considered the even more difficult  problem of constructing data with a complete metric on 
$\mathbb{R}^3$ that are exactly  Schwarzschild in a neighbourhood of space-like infinity. 
They managed to construct  for the Einstein-Maxwell equations  a smooth family of such data which includes the Minkowski data \cite{cutler:wald}. 
This allowed them to show for the first time the {\it existence non-trivial solutions to the 
 Einstein-Maxwell equations which are null geodesically complete and admit smooth conformal extensions at null infinity with complete ${\cal J}^{\pm}$ and regular points $i^{\pm}$}. In fact, 
 the control on the asymptotic behaviour of the solutions in the static  region near space-like infinity allows them to conclude that the solutions contain hyperboloidal hypersurfaces with induced data that can be arbitrarily close to Minkowskian hyperboloidal data. Invoking the strong stability result on Minkowskian  hyperboloidal developments then gives the result.

The data so constructed look highly special but the work initiated by Corvino shows that there exist in fact large classes of data with specialized ends.

\vspace{.1cm}

\noindent
{\bf Complete vacuum data with Schwarzschild  ends.}  \cite{corvino:2000}: { \it A given time reflection symmetric, asymptotically  flat vacuum data set can be deformed outside some prescribed compact domain so as to obtain  vacuum data which are exactly Schwarzschild in some neighbourhood of spatial infinity.}

\vspace{.1cm}

This has been generalized by Corvino and Schoen  (\cite{corvino:schoen})  and Chru\'sciel and Delay (\cite{chrusciel:delay:2003}), who  obtained vacuum data which agree on prescribed compact sets
 with given asymptotically flat data and which are stationary near spatial infinity. 
 (It follows in particular that in contrast to static or stationary data, which are  determined completely
 by  their  multipoles at space-like infinity, such quantities are of little significance in the context of more general data, where they  hardly carry relevant information about the structure of the data in the interior.) This work  also led to  generalizations of the Cutler - Wald idea.
  
 \vspace{.1cm}

\noindent
{\bf Vacuum solutions with smooth asymptotic structure} \cite{corvino: 2007}, 
\cite{chrusciel:delay:2002}:
{\it There exist large classes of vacuum solutions that are null geodesically complete
and static or stationary near space-like infinity, which admit smooth and complete conformal 
boundaries ${\cal J}^{\pm}$ at null infinity and regular points $i^{\pm}$ at future and past time-like infinity.}

\vspace{.5cm}

It appears that we are getting close to obtaining conditions which are {\it necessary and sufficient 
for asymptotically flat solutions to admit a smooth conformal extension in a neighborhood of the critical set} or, with suitable smallness assumptions on the data, along complete null infinities. 
But closing the gap still requires some complicated analysis. There are good arguments why it would be worth the effort.

In the time reflection symmetric case  it has been shown in \cite{friedrich:kannar1} that, given enough smoothness, the completion of the picture reduces many questions about the asymptotics and the associated physical concepts which have been discussed in the literature for a long time  to straightforward (though possibly lengthy) calculations. In particular, the settings of \cite{newman:penrose}, \cite {newman:penrose:NP2} are related to the setting of \cite{friedrich:i-null} near space-like infinity,  it is shown how quantities of physical interest like the 
Bondi-mass or the NP conserved quantities near the critical set on ${\cal J}^+$ can be related explicitly to quantities on the initial slice near spatial infinity like the ADM-mass and higher order expansion coefficients etc. 
The setting also allows us to single out in a unique  way the Poincare  group as a subgroup of  the BMS-group. It would be interesting to understand the weakest smoothness conditions under  which this can still be done and under which the  BMS-group can still be defined.
Some generalizations of these results are discussed in \cite{j.a.v.kroon:2007}.

The precise identification of the data content  which needs to be supressed to achieve a certain amount of asymptotic smoothness may help understand the role and meaning of that content (if there is any). Referring to it as `radiation near space-like infinity' explains very little. It needs to be decided whether there exist physical systems of interest which require this content for their adequate modelling.

The setting proposed in \cite{friedrich:i-null} offers the possibility to calculate numerically entire 
asymptotically flat solutions,  including their asymptotics and radiation fields, on finite grids. 
First steps to develop adequate numerical methods  are being  taken in \cite{beyer:doulis:frauendiener:whale:2012}
and \cite{frauendiener:hennig:2014}. The work in \cite{beyer:doulis:frauendiener:whale:2012} 
even includes a discussion of the logarithmic terms exhibited  in \cite{friedrich:spin-2}
and it suggests that the difficulties arising from the loss of hyperbolicity of the equation at the critical sets can be compensated by the information supplied by the transport equations on the cylinder $I$ at space-like infinity.

An important open problem is the numerical calculation of Corvino-type data. 
In the time reflection symmetric case there has been proposed a general method  how to obtain such data by solving PDE systems \cite{avila:2011}. But this needs further analytical and numerical study.
If such data can be provided numerically, it will help analyzing the effect of Corvino-type deformations on the developments of solutions  in time, a subject of greatest  interest from our point of view. 
In the conformal picture the solutions seem to be modified  by such deformations only  in some `thin' neighbourhood of ${\cal J}^- \cup i^0 \cup {\cal J}^+$. But to what extent do they affect the structure of the radiation field in the causal  future  of that part of the initial hypersurface where the essential physical processes take place ?
What is the difference in the radiation fields resulting from two different  Corvino-type deformations ?
If the radiation field on ${\cal J}^+$ in the causal future of a merger process
would hardly be affected by deformations that are performed sufficiently close to space-like infinity, there would be no reason to worry about the data asymptotics. If it would be affected in an essential way, however, choosing the `correct' asymptotics of the Cauchy data would become a delicate matter in any case.

\section{Anti-de Sitter-type solutions.}
\label{negative lambda}

The AdS/CFT correspondence proposed by Maldacena \cite{maldacena:1998} triggered an enormous interest in solutions to Einstein's field equations with cosmological constant $\lambda < 0$. 
There exists, however, no observational evidence which would motivate the choice $\lambda < 0$ physically 
and the analysis must be guided by what looks mathematically natural and reasonably general.
We shall only consider  four-dimensional solutions with $\lambda < 0$
and view them, like de Sitter- and Minkowski-type solutions,
as classical relativistic objects representing cosmological models or subsystems thereof.

The model to be generalized in the following is the simply connected, conformally flat 
{\it anti-de Sitter covering space}, short AdS, which is given by

\vspace{.1cm}

\hspace*{2cm}$\hat{M} = \mathbb{R} \times \mathbb{R}^{3}, \quad \quad
\hat{g} =  - \cosh^2 r\,dt^2 + dr^2 + \sinh^2 r\,h_{\mathbb{S}^{2}}$,

\vspace{.1cm}

\noindent
where $r \ge 0$ denotes a radial coordinate on $ \mathbb{R}^{3}$. It solves (\ref{vac-lambda}) with $\lambda = - 3$.
A clear picture of its global and asymptotic structure  is obtained by 
combining the coordinate transformation  $\rho = 2\,\arctan(e^r) - \frac{\pi}{2}$ with a rescaling by the conformal factor $\Omega = \cosh^{-1} r = \cos \rho$ to obtain the conformal representation

\vspace{.1cm}

\hspace*{2cm}$g = \Omega^2\,\hat{g} = - dt^2 + d\rho^2 + \sin^2\rho\,\,h_{\mathbb{S}^{2}},\quad 
\quad 0 \le \rho < \frac{\pi}{2}$.

\vspace{.1cm}

 \noindent
The metric $g$ and the conformal factor $\Omega$ extend smoothly as $\rho \rightarrow \frac{\pi}{2}$ and then live on the manifold $M = \mathbb{R} \times  \overline{\mathbb{S}^{3}/2}$ where the second factor denotes the closure of a hemisphere of $\mathbb{S}^{3}$. The boundary ${\cal J} = \{\rho = \frac{\pi}{2}\} \sim  \mathbb{R} \times \mathbb{S}^{2}$ attached by this process to $\hat{M}$ is time-like for $g$ and its  points can be understood  as endpoints of the space-like and null geodesics of $\hat{g}$ so that ${\cal J}$  represents space-like and null infinity for AdS.

Any solution of Einstein's equations (\ref{vac-lambda}) with $\lambda < 0$  which admits in a similar way a smooth conformal extension that  adds a time-like hypersurface ${\cal J}$ which represents space-like and null infinity will be referred to as an {\it  AdS-type space-time.} Two basic features distinguish the  global  causal resp. conformal structure of AdS from that of de Sitter- or Minkowski-space.  The obvious one is the fact that AdS (in fact any AdS-type space-time) fails to be globally hyperbolic. A less obvious one will be discussed in the context of the AdS stability problem, where it will become as important as the presence of a time-like boundary.

If more general AdS-type solutions to  (\ref{vac-lambda}) are to be constructed by solving PDE  problems, one needs to analyze the  freedom to prescribe boundary data. The formal expansions at the the conformal boundary found in the literature amount  to analyzing Cauchy problems with data on ${\cal J}$. This may be of interest in some contexts  but will not suffice for us. Cauchy problems for hyperbolic equations with data on time-like hypersurfaces are known to be ill-posed. Moreover, if the formal expansions can be shown to define real analytic solutions in some neighborhood of ${\cal J}$ (see \cite{kichenassamy:2004}) the analytic extension into the interior will, more likely than not, end in a singularity and it remains unclear whether these solutions admit any extension at all which is regular in the sense that it contains a complete (in the induced metric) space-like hypersurface that intersects ${\cal J}$ in a space-like surface. 

\subsection{An existence result}

The natural problem to consider  is the initial-boundary value problem with boundary data prescribed on ${\cal J}$  and Cauchy data given on a space-like slice that extends to the boundary.

\vspace{.1cm}

\noindent
{\bf Existence of AdS-type soutions local in time} \cite{friedrich:AdS}:
{\it Suppose $\lambda$ is a negative number and $(\hat{S}, \hat{h}_{ab}, \hat{\chi}_{ab})$ is a smooth Cauchy  data set for (\ref{vac-lambda}) with $\hat{\kappa} = \hat{h}^{ab}\, \hat{\chi}_{ab} = const. \neq 0$ 
so that  $\hat{S}$ is an orientable $3$-manifold and
$(\hat{S}, \hat{h}_{ab})$ is a complete Riemannian space. Let these data  admit a smooth conformal completion 

\vspace{.1cm}

\hspace*{1cm}$\hat{S} \rightarrow S = \hat{S} \cup \Sigma, \quad
\hat{h}_{ab} \rightarrow h_{ab} = \Omega^2\,\hat{h}_{ab}, \quad
\hat{\chi}_{ab} \rightarrow \chi_{ab} = \Omega\,(\hat{\chi}_{ab} - \frac{1}{3}\,\hat{\kappa}\,\hat{h}_{ab})$,

\vspace{.1cm}

\noindent
where
 $(S, h_{ab})$ is a Riemannian space with compact boundary $\partial S = \Sigma$, 
$\Omega$ a smooth defining function of $\Sigma$ with $\Omega > 0$ on $ \hat{S}$,
and  
$W^{\mu}\,_{\nu \lambda \rho} = \Omega^{-1}\, \hat{C}^{\mu}\,_{\nu \lambda \rho}$ 
extends smoothly to $\Sigma$ on $S$ where $ \hat{C}^{\mu}\,_{\nu \lambda \rho}$ denotes then conformal Weyl tensor determined by the metric $ \hat{h}_{ab}$ and the second fundamental form $ \hat{\chi}_{ab}$.

Consider the boundary ${\cal J} = \mathbb{R} \times \partial S$ of $M = \mathbb{R} \times S$ 
and  identify 
$S$ with $\{0\} \times S \subset M$ and
$\Sigma$ with $\{0\} \times \partial S = S \cap {\cal J}$.
Let on ${\cal J}$  be given a smooth 3-dimensional  Lorentzian
conformal structure which satisfies in an adapted gauge together with the Cauchy data  the {\it corner conditions} implied on $\Sigma$  by the conformal field equations, where it is  assumed that the normals to $S$ are tangent to ${\cal J}$ on  $\Sigma$.

Then there exists for some $t_o > 0$ on the set  
$\hat{W} = \, ]- t_o,   t_o[ \times \hat{S} \subset \mathbb{R} \times \hat{S} \subset M$
a unique solution $\hat{g}$ to  (\ref{vac-lambda})  which  admits 
with some smooth boundary defining function $\Omega$ on $M$ (that extends the function 
$\Omega$ on $S$ above) a smooth conformal extension 

\vspace{.1cm}

\hspace*{3cm}$\hat{W}  \rightarrow W =   ]-   t_o,   t_o[ \times S,
\quad \hat{g}  \rightarrow g = \Omega^2\hat{g}$,

\vspace{.1cm}

 \noindent
that induces  (up to a conformal diffeomorphism) on $S$ and 
${\cal J}_o = ]-   t_o,   t_o[ \times \partial S$  the given conformal data.}

\vspace{.5cm}

This is  the first and still the only well-posed initial boundary value problem for Einstein's field equations which is general in the sense that no symmetries are required and which admits a covariant formulation  
\cite{friedrich:geom-unique}. 
It supports the view that the setting of asymptotic simplicity is natural for Einstein's equations. The idea of a smooth conformal extension is basic for the formulation of the PDE problem which leads to this result. All possible AdS-type solutions local in time (with $|t_o|$ sufficiently small) are obtained. 
In the following we point out the particular features of AdS-type vacuum solutions which allow one to obtain this result, and comment on a particular choice of boundary condition. 

In \cite{friedrich:AdS}  has been observed a correspondence between Cauchy data sets as required above and  hyperboloidal Cauchy data for Einstein's equations (\ref{vac-lambda}) with $\lambda = 0$.
This has been worked out in detail in \cite{kannar:1996b}.  It is not known
whether the existence result local in time can be extended to hyperboloidal data which only admit poly-homogeneous expansions at infinity. Most likely the non-smoothness will spread into the physical manifold.

Because the data on ${\cal J}$ are not subject to constraints the discussion of the boundary conditions looks simple. The covariant formulation rests, however, on specific properties of AdS-type solutions and is obtained in two steps. In the first step boundary conditions/data are considered which relate directly to a well-posed PDE problem.
AdS-type solutions have the special feature that the second fundamental form $\kappa_{ab}$ induced on ${\cal J}$ 
can be made to vanish on ${\cal J}$ in a suitable conformal  gauge.
As a consequence there exists a certain  geometric gauge in which the location of the boundary ${\cal J}$ is known, 
the conformal factor $\Omega$ with $\Omega = 0$ and $d\Omega \neq 0$  on ${\cal J}$ is known explicitly, and on ${\cal J}$ the gauge is determined  in terms of the inner metric induced on 
${\cal J}$. 
In this gauge the conformal field equations  imply hyperbolic evolution equations which assume in Newman-Penrose notation the form

\vspace{.1cm}

\hspace*{1.2cm}$\partial_{\tau}u = F(u, \psi, x^{\mu})$,
\hspace*{1cm}$(1 + A^0)\,\partial_{\tau}\psi + A^{\,\alpha}\,\,\partial_{\alpha}\psi = G(u, \psi, x^{\mu})$,

\vspace{.1cm}

 \noindent
where $\tau = x^0$ is a time coordinate, $x^{\alpha}$ are spatial coordinates, and the unknown $u$ comprises besides the 
pseudo-orthonormal  frame coefficients 
$(e^{\mu}\,_k )_{k = 0, .., 3} $ also the connection coefficients  and the Schouten tensor 
$L_{jk} = \frac{1}{2}\,(R_{jk} - \frac{1}{6}\,R\,g_{jk}) $ of the metric $g$ with respect to this frame. The matrices $A^{\mu}$ depend on the frame coefficients and the coordinates and $\psi = (\psi_0, \ldots \psi_4)$  represents the essential components of the symmetric spinor field $\psi_{ABCD}$ that corresponds to the tensor field 
$W^{i}\,_{jkl}$. The frame is chosen so that the future directed time-like vector field $e_0 + e_1$ is tangential to 
${\cal J}$ and the space-like vector field $e_0 - e_1$ is normal to ${\cal J}$ and inward pointing. Then $e_2$, $e_3$ are tangential to ${\cal J}$, $e_0$ is inward  and $e_1$ is outward pointing on ${\cal J}$. General results on initial-boundary value problems then give boundary conditions of the form

\vspace{.05cm}

\hspace*{2.5cm}$\psi_{4} - a\,\psi_{0} - c\,\bar{\psi}_{0} = d  = d_1 + i\,d_2, \quad |a| + |c| \le 1\,\,$  on $\,\,{\cal J}$, 

\vspace{.1cm}

 \noindent
where the smooth complex-valued function $d$ on ${\cal J}$ denotes the essential free boundary datum and the smooth complex-valued functions $a$ and $c$  on ${\cal J}$ can be chosen freely within the indicated restrictions.

Given the conformal Cauchy data on $S$ in the  gauge used above, the formal expansion of the unknowns $u$, 
$\psi$  in terms of the coordinate $\tau$ is determined at all orders on $S$,   in particular on $\Sigma$. 
On ${\cal J}$ let be given smooth functions $a$ and $c$ as above. Using their formal expansion in terms of $\tau$ on $\Sigma$, the formal expansion of the term on the left  hand side of of the boundary condition on $\Sigma$ is obtained. The corner conditions consist in the requirement that this formal expansion coincides on $\Sigma$  with the formal expansion of the free boundary datum $d$. Borel's theorem guarantees that there always exist smooth functions $d$ on ${\cal J}$ which satisfy this requirement. Away from $\Sigma$ they are essentially arbitrary.

\vspace{.1cm}

With Cauchy data as stated in the theorem and boundary conditions as above
where $d$ satisfies the corner conditions one obtains a {\it well-posed initial-boundary value problem which preserves the constraints and the gauge conditions}. This implies the  existence and uniqueness of smooth solutions on a domain as indicated in the theorem.

\vspace{.1cm}

The formulation so obtained has the drawback that the boundary condition  depends  implicitly on the choice of of the time-like vector field n$e_0 + e_1$. This reflects  a general difficulty with initial boundary-value problems for Einstein's  equations \cite{friedrich:geom-unique}.  In the case of  AdS-type solutions it can be overcome by using the observation that such solutions must satisfy the relation $w^*_{ab}  = \sqrt{3/|\lambda|}\,B_{ab}$ on ${\cal J}$,
where $w^*_{ab}$  denotes the ${\cal J}$-magnetic part of  $W^{i}\,_{jkl}$, obtained be contracting  the right dual of $W^{i}\,_{jkl}$ twice with the inward pointing unit normal of ${\cal J}$, and  $B_{ab}$  is the  (dualized) Cotton tensor of the metric $k_{ab}$ induced on  ${\cal J}$. With the particular choice $\psi_{4} - \,\bar{\psi}_{0} = d_1 + i\,d_2$ of the boundary condition this allows one to express certain components of the Cotton tensor in terms of the real-valued functions $d_1$, $d_2$. If these components are given, the structural equations of the normal conformal Cartan connection defined by the conformal structure on ${\cal J}$ provides in our gauge  a hyperbolic differential system  on ${\cal J}$ which determines, with the data given on $\Sigma$,  the inner metric $k_{ab}$ uniquely in terms of  $d_1$ and $d_2$. Conversely, the function $\psi_{4} - \,\bar{\psi}_{0}$, whence the free data $d$, can be calculated in the given gauge  uniquely from the inner metric $k_{ab}$ on ${\cal J}$.

\vspace{.1cm}

Irrespective of the functions $a$ and $c$, 
any boundary condition with $d = 0$ on ${\cal J}$
can be considered as a reflective boundary condition. With an additional requirement on 
$\Sigma = S \cap {\cal J}$ it implies the relation $B_{ab}[k] = 0$ of local conformal flatness on ${\cal J}$.  This will be referred to as the {\it reflective boundary condition}.
It appears so natural to many authors that they refer to it as  {\it the AdS-boundary condition}. 
There seems to be no particular reason, however,  why it should be preferred. 
Moreover, when  conditions are imposed  on the boundary data, the clean separation between the evolution problem and the problem of the constraints we are used to from the standard Cauchy problem, is not  maintained any longer. 
{\it If reflective boundary conditions are imposed on ${\cal J}$, 
consistency with the corner conditions requires that
the Cauchy data on $S$ satisfy besides the underdetermined elliptic constraint equations on $S$ an infinite number of differential conditions at $\Sigma$}  \cite{friedrich:ADS-stabiity}.

\vspace{.1cm}

The results obtained in four space-time dimensions  in the cases 
$\lambda \ge 0$ have been generalized to even space-time dimensions larger than four. It is conceivable that 
the type of analysis of the initial boundary value problem for Einstein's field equations
in  \cite{kreiss:et:al:2009} can be applied to
the conformal differential system considered in \cite{anderson:2005a}
to obtain such a generalization also in the case $\lambda < 0$. This requires, however, a new study of the problem of constraint propagation and  different arguments to obtain covariant boundary conditions.

\subsection{On the stability problem.}

Given the local existence result, there arises the question whether the solution can be controlled globally in time or, more modestly,  whether AdS is non-linearly stable. It turns out that this stability problem is more challenging  than the corresponding ones in the cases $\lambda \ge 0$. There is the technical difficulty of controlling the evolution for an arbitrary length of time, but it is as already  problematic to say what should be meant by `stability' in the present context.

Bizo\'n and Rostworowski \cite{bizon:rostworowski:2011} recently presented a first study of the  AdS-stability 
problem by using  mainly numerical methods.  Their work raises some extremely interesting questions concerning solutions to Einstein's equations  (\ref{vac-lambda}) with $\lambda < 0$  that are subject to conditions on the boundary ${\cal J}$ at space-like and null infinity. They analyse  the spherically symmetric Einstein-massless-scalar field  system with homogeneous Dirichlet asymptotics and Gaussian type initial data  and observe the formation of trapped  surfaces for (numerically)  arbitrarily small  initial data.  They supply  numerical evidence that the development of trapped surfaces results  from an energy transfer from low to high frequency modes.  They perform a perturbative analysis, which points into the same direction but also exhibits   {\it small one-mode initial data} which develop into forever smooth solutions. It is not so clear whether there  are also small neighborhoods of these data which develop into (numerically) forever  smooth solutions.  The results led them to  suggest: {\it AdS is unstable against the formation of black holes for a large class of arbitrarily small perturbations}. 

While it has been formulated in the context of a particular model this conjecture may easily be understood as applying more generally. In the following we wish to point out that the situation considered here is 
extremely special and may obscure the view onto a scenery which is much richer in possibilities.

The setting considered in \cite{bizon:rostworowski:2011}, mainly motivated by questions of technical 
feasibility, involves reflecting boundary conditions.  These are convenient because they lead to clean initial boundary value problems but they are also very restrictive from a physical point of view. Such  systems cannot  interact with an ambient universe  and thus certainly  do not represent observable objects as suggested by some of the names  given to them in the literature.

For the stability problem a second feature of the global conformal structure of AdS is of equal importance  as the existence of the time-like boundary. Contrary to what is occasionally suggested in the literature and what does hold in the cases  $\lambda \ge 0$, AdS does not admit conformal rescalings that put past or future time-like infinity, represented by points or more general sets, in a  finite coordinate  location {\it and extend smoothly}.  In this sense AdS is always of infinite length in time, even in `conformal time' (this can be supported by rigorous arguments).

Some of the observations in  \cite{bizon:rostworowski:2011} may then not be too surprising. By the non-linearity of the Einstein equations the spherically symmetric field perturbation can be expected to experience a focussing effect when it  travels through the center and the  boundary condition  leads to a reflection and refocussing of the perturbation at the boundary.   Because conformal time is potentially unlimited, this process can repeat itself arbitrarily often and the focussing effects, however tiny at each step, may eventually add up to produce a collapse.
The occurrence of  islands of stability then appears in fact more surprising  
than the tendency to develop a collapse.

The relatives sizes of the ocean of data of instability and the tiny  islands of stability may change drastically  if the full freedom to impose boundary conditions is taken into account. As long as physical considerations do not tell us what kind of objects should be represented by
solutions with $\lambda < 0$ it does not appear advisable to exclude any choices  in the stability analysis.

It is instructive to compare the different situations. In the case $\lambda > 0$ there is no ambiguity in saying that initial data on a (compact) Cauchy hypersurface are close to some reference data like de Sitter data. Incoming information can be specified on ${\cal J}^-$ and outgoing  information can be controlled on ${\cal J}^+$.  
If the information entering at ${\cal J}^-$ is trivial, the solution is trivial. Potential difficulties arising from spatial compactness are compensated by the exponential expansion of the solutions.

If initial data are prescribed in the case $\lambda = 0$ on a Cauchy  hypersurface, the requirement that they be close to Minkowskian data leaves ambiguities near space-like infinity which may affect  the smoothness at 
${\cal J}^{\pm}_*$. But under fairly general  assumptions  we find again that there is a clear separation between incoming information, specified in terms of the radiation field on ${\cal J}^-$, and outgoing  information, registered by the radiation field on  ${\cal J}^+$. If the  incoming information is trivial the solution is trivial.

In the AdS-type case $\lambda < 0$ there exists a data  ambiguity at the spatial end of the initial slice and a related ambiguity all along the boundary ${\cal J}$.
Refusing  information to enter the space-time through ${\cal J}$ by imposing reflecting boundary conditions does not lead to trivial solutions. There do exist non-trivial  vacuum initial data consistent with these boundary conditions \cite{chrusciel:delay:2009}. Moreover, under more general assumptions on the boundary conditions/data it is far from obvious how
to control in- and outgoing radiation separately  to achieve a balance along ${\cal J}$ which avoids the development of a gravitational collapse.

}


\begin{thebibliography}{9}




\bibitem{acena:kroon}
A. E. Ace\~na, J. A. Valiente Kroon. 
\newblock Conformal extensions for stationary space-times.
\newblock {\em Class. Quantum Grav.} 28 (2011) 225023 



\bibitem{anderson:2005a}
M. Anderson
\newblock Existence and stability of even-dimensional asymptotically de Sitter spaces.
\newblock {\em Ann. Henri Poincar\'e} 6 (2005) 801 - 820. 



\bibitem{anderson:chrusciel:2005}
M. Anderson, P. Chru\'sciel.
\newblock Asymptotically simple solutions of the vacuum Einstein equations in even dimensions.
\newblock {\em Commun. Math. Phys.} 260 (2005) 557 - 577.



\bibitem{andersson:chrusciel:1996}
L. Andersson, P. Chru\'sciel.
\newblock Solutions of the constraint equations in general relativity satisfying `hyperboloidal boundary conditions'.
\newblock {\em Dissertationes Mathematicae}, Polska Akademia Nauk, Inst. Matem.,
Warszawa, 1996.

\bibitem{andersson:chrusciel:friedrich:1992}
L. Andersson, P. Chru\'sciel, H. Friedrich.
\newblock On the existence of solutions to the Yamabe equation and the existence of smooth  hyperboloidal initial data for Einstein's field equations.
\newblock  {\em Commun. Math. Phys.} 149 (1992) 587 - 612.

\bibitem{andersson:galloway:2002}
L. Andersson, G. Galloway.
\newblock dS/CFT and space-time topology.
\newblock   {\em Adv. Theor. Math. Phys.} 6 (2002) 307 - 327. 

\bibitem{avila:2011}
G. Avila.
\newblock Asymptotic staticity and tensor decompositions with fast decay conditions.
\newblock Thesis, University of Potsdam, 2011.
\newblock http://opus.kobv.de/ubp/volltexte/2011/5404/ 


\bibitem{beig:simon1980}
R. Beig, W. Simon.
\newblock Proof of a multipole conjecture due to Geroch.
\newblock {\em Comm. Math. Phys.} 78 (1980) 75 - 82.

\bibitem{beig:simon1981}
R. Beig, W. Simon.
\newblock On the multipole expansion for stationary space-times..
\newblock {\em Proc. R. Soc.} Lond. A 376 (1981) 333-341.

\bibitem{beyer:2007}
F. Beyer.
\newblock Asymptotics and singularities in cosmological models with positive cosmological constant.
\newblock http://arxiv.org/abs/0710.4297



\bibitem{beyer:2008}
F. Beyer.
\newblock Investigations of solutions of Einstein's field equations close to $\lambda$-Taub-NUT.
\newblock   {\em Class. Quantum Grav.} 25 (2008) 235005.

\bibitem{beyer:2009a}
F. Beyer.
\newblock Non-genericity of the Nariai solutions. I Asymptotics and spatially homogeneous perturbations.
\newblock   {\em Class. Quantum Grav.} 26 (2009) 235015.

\bibitem{beyer:doulis:frauendiener:whale:2012}
F. Beyer, G. Doulis, J. Frauendiener, B. Whale.
\newblock Numerical space-times near space-like and null infinity: The spin-2-system on Minkowski space.
\newblock  {\em Class. Quantum Grav.} 29 (2012) 245013..



\bibitem{bicak}
J. Bi$\check{c}$\'ak.
\newblock Selected solutions of Einstein's field equations: Their role in general relativity and astrophysics.
\newblock In: B.Schmidt (ed.) {\it Einstein's field equations and their physical implications}.
\newblock Springer, Berlin, 2000.






\bibitem{bieri:zipser:2009}
L. Bieri, N. Zipser.
\newblock {\em Extensions of the stability theorem of the Minkowski space in general relativity}.
\newblock   AMS, International Press, 2009. 

\bibitem{bieri:2009}
L. Bieri.
\newblock Priviate communication (2012).






\bibitem{bizon:mach:2014}
P. Bizo\'n, P. Mach.
\newblock Global dynamics of a Yang-Mills field on an asymptotically hyperbolic space.
\newblock arXiv: 1410.4317 [math.AP]



\bibitem{bizon:rostworowski:2011}
P. Bizo\'n, A. Rostworowski.
\newblock  On weakly turbulent stability of anti-de Sitter spacetime.
\newblock   {\em Phys. Rev. Lett.} 107 (2011) 031102. 


\bibitem{bondi:et.al}
H. Bondi, M. G. J. van der Burg, A. W. K. Metzner.
\newblock Gravitational waves in general relativity VII. Waves from
axi-symmetric isolated systems.
\newblock { Proc. Roy. Soc} A 269 (1962) 21--52.


\bibitem{choptuik}
M. W. Choptuik.
\newblock Universality and scaling in gravitational collapse of a
massless scalar field.
\newblock {\em Phys. Rev. Lett.} 70 (1993) 9.



\bibitem{choquet-bruhat:1952}
Y. Choquet-Bruhat.
\newblock Th\'eor\`emes d'existence pour certains syst\`emes 
d'\'equations aux de\'riv\'ees
partielles non lin\'eaires.
\newblock {\em Acta. Math.} 88 (1952) 141 - 225.


\bibitem{choquet-bruhat:geroch:1969}
Y. Choquet-Bruhat, R. Geroch.
\newblock Global aspects of the Cauchy problem in general relativity.
\newblock {\em Commun. Math. Phys.} 14 (1969) 329 - 335.


\bibitem{choquet-bruhat:novello:1987}
Y. Choquet-Bruhat, M. Novello.
\newblock Syst\`eme conforme r\'egulier pour les \'equations d'Einstein.
\newblock {\it C. R. Acad. Sci. Paris} 305 (1987) 155 -160.




\bibitem{christodoulou:klainerman:1993}
D. Christodoulou, S. Klainerman.
\newblock {\em The global non-linear stability of Minkowski space}.
\newblock Princeton University Press, Princeton, 1993.



\bibitem{chrusciel-maccallum-singleton:1995}
P. T. Chru\'sciel, M. A. H. Mac Callum, D. B. Singleton.
\newblock  Gravitational waves in general relativity XIV. Bondi expansion and the 
`polyhomogeneity' of ${\cal J}$.





\bibitem{chrusciel:delay:2002}
P. T. Chru\'sciel, E. Delay.
\newblock Existence of non-trivial, vacuum, asymptotically simple
spacetimes.
\newblock {\em Class. Quantum Grav.}, 19 (2002) L 71 - L 79.
\newblock Erratum
\newblock {\em Class. Quantum Grav.}, 19 (2002) 3389.

\bibitem{chrusciel:delay:2003}
P. T. Chru\'sciel, E. Delay.
\newblock  On mapping properties of the general relativistic constraints operator in weighted 
function spaces, with application. 
\newblock {\it M\'em. Soc. Math. Fr. (N.S.)} 94 (2003).


\bibitem{chrusciel:delay:2009}
P. T. Chru\'sciel, E. Delay.
\newblock Gluing constructions for asymptotically hyperbolic manifolds with constant scalar curvature.
\newblock {\it Communications in Analysis and Geometry} 17 (2009) 343 - 381.


\bibitem{chrusciel-paetz-2013}
P. T. Chru\'sciel, T.-T. Paetz.
\newblock Solutions of the vacuum Einstein equations with initial data on past null infinity.
\newblock {\it Class. Quantum Grav.} 30 (2013) 235037.




\bibitem{corvino:2000}
J. Corvino.
\newblock Scalar curvature deformation and a gluing
construction for the Einstein constraint equations.
\newblock { Comm. Math. Phys.} 214 (2000) 137--189. 

\bibitem{corvino: 2007}
J. Corvino.
\newblock On the existence and stability of the Penrose compactification.
\newblock {\it Ann. Henri Poincar\'e} 8 (2007) 597 - 620.

\bibitem{corvino:schoen}
J. Corvino, R. Schoen
\newblock On the Asymptotics for the Vacuum Einstein Constraint
Equations.
\newblock  {\it J. Differential Geometry}  73 (2006) 185 - 217.


\bibitem{couch-torrence:1972}
W. E. Couch, R. J. Torrence.
\newblock Asymptotic behaviour of vacuum space-times.
\newblock {\it J. Math. Phys.} 13 (1972) 69 - 73.


\bibitem{cutler:wald}
C. Cutler, R. M. Wald.
\newblock Existence of radiating Einstein-Maxwell solutions which are $C^{\infty}$ on all of 
${\cal J}^+$ and ${\cal J}^-$.
\newblock {\it Class. Quantum Grav.}  6 (1989) 453 - 466.

\bibitem{dain:stationary}
S. Dain.
\newblock Initial data for stationary spacetimes near spacelike infinity.
\newblock {\em Class. Quantum Grav.}, 18 (2001) 4329--4338.

\bibitem{dain:friedrich}
S. Dain, H. Friedrich.
\newblock Asymptotically Flat Initial Data with
Prescribed Regularity.
\newblock {\it  Comm. Math. Phys.}  222 (2001) 569--609. 

\bibitem{davis:2014}
T. M. Davis.
\newblock Cosmological constraints on dark energy.
\newblock {\it Gen. Relativ. Gravit.} 46 (2014) 1731



\bibitem{frauendiener:2004}
J. Frauendiener.
\newblock Conformal infinity.
\newblock Living Rev. Relativity 7 (2004)
\newblock http://www.livingreviews.org/lrr-2004-1 
              
      
\bibitem{frauendiener:hennig:2014}
J. Frauendiener, J. Hennig.
\newblock  Fully pseudospectral solution of the conformally invariant wave equations near the cylinder at space-like infinity.      
\newblock {\em Class. Quantum Grav.}, 31 (2014) 085010.
                   
\bibitem{friedrich:1981a}
H. Friedrich.
\newblock On the regular and the asymptotic characteristic initial value
problem for Einstein's vacuum field equations. 
\newblock Proceedings of the 3rd Gregynog Relativity
Workshop on  Gravitational Radiation Theory
\newblock MPI-PEA/Astro 204 (1979) 137-160 and
\newblock { Proc. Roy. Soc.}, 375 (1981) 169-184.

\bibitem{friedrich:asymp-sym-hyp:1981b}
H.  Friedrich. 
\newblock  The asymptotic characteristic initial value problem for
Einstein's vacuum field equations as an initial value problem
for a first-order quasilinear symmetric hyperbolic
system.
\newblock  {\it Proc.\ Roy.\ Soc.\ Lond.\ A } 378 (1981) 401-421. 

\bibitem{friedrich:1982}
H. Friedrich.
\newblock On the Existence of Analytic Null Asymptotically Flat
Solutions of Einstein's Vacuum Field Equations.
\newblock { Proc. Roy. Soc.\ Lond.\ A} 381 (1982) 361 - 371.

\bibitem{friedrich:1983}
H. Friedrich.
\newblock Cauchy problems for the conformal vacuum field equations
in  General Relativity.
\newblock {\it Commun.\ Math.\ Phys.} 91 (1983) 445-472.


\bibitem{friedrich:hyperbolic}
H. Friedrich.
\newblock On the hyperbolicity of Einstein's and other gauge field equations.
\newblock {\em Commun. Math. Phys.}, 100 (1985) 525 - 543.


\bibitem{friedrich:1986a}
H. Friedrich.
\newblock Existence and structure of past asymptotically simple solution of 
Einstein's field equations with positive cosmological constant.
\newblock {\it J. Geom. Phys.}, (1986) 101 - 117.


\bibitem{friedrich:1986b}
H. Friedrich.
\newblock On the existence of n-geodesically complete or future 
complete solutions of Einstein's field equations with smooth asymptotic structure.
\newblock {\it Commun. Math. Phys.} 107 (1986), 587 - 609.


\bibitem{friedrich:pure rad}
H. Friedrich.
\newblock On purely radiative space-times.
\newblock {\em Commun. Math. Phys.} 103 (1986) 35 - 65. 


\bibitem{friedrich:1991}
H. Friedrich.
\newblock On the global existence and the asymptotic behaviour of solutions
 to the Einstein-Maxwell-Yang-Mills equations.
\newblock {\it J. Differential Geometry} 34 (1991) 275 - 345.

\bibitem{friedrich:AdS}
H. Friedrich.
\newblock Einstein equations and conformal structure: existence of
anti-de Sitter-type space-times.
\newblock { \it J. Geom. Phys.}, 17 (1995) 125--184.


\bibitem{friedrich:2hyp red}
H. Friedrich.
\newblock Hyperbolic Reductions for Einstein's Equations.
\newblock { Class. Quantum Gravity} 13 (1996) 1451--1469.

\bibitem{friedrich:i-null}
H. Friedrich.
\newblock Gravitational fields near space-like and null
infinity.
\newblock { \it J. Geom. Phys.}  24 (1998)  83--163.

\bibitem{friedrich:tueb}
H. Friedrich.
\newblock Conformal Einstein evolution.
\newblock In: J. Frauendiener, H. Friedrich (eds.) {\it The Conformal Structure of
Spacetime:  Geometry, Analysis, Numerics.} Springer, Berlin, 2002.


\bibitem{friedrich:spin-2}
H. Friedrich.
\newblock Spin-2 fields on Minkowski space near spacelike and
null infinity.
\newblock {\em Class. Quantum. Grav.} 20 (2003) 101 - 117.


\bibitem{friedrich:cargese}
H. Friedrich.
\newblock Smoothness at null infinity and the structure of initial data.
\newblock In: P. T. Chru\'sciel, H. Friedrich (eds.):
{\it The Einstein Equations and the Large Scale Behaviour of Gravitational Fields.}
\newblock Birkh\"auser, Basel, 2004.

\bibitem{friedrich:DPG}
H. Friedrich.
\newblock Is general relativity `essentially understood' ?
\newblock  {\em Ann. Phys. (Leipzig)} 15 (2006) 84 - 108.

\bibitem{friedrich:statconv}
H. Friedrich.
\newblock Static vacuum solutions
from convergent null data expansions at space-like infinity.
\newblock {\em Ann. Henri Poincare} 8 (2007) 817 - 884.

\bibitem{friedrich:geom-unique}
H. Friedrich.
\newblock Initial boundary value problem for Einstein's field equations and geometric uniqueness.
\newblock {\it Gen Relativ Gravit} 41 (2009) 1947 - 1966.

\bibitem{friedrich:conf-str-static-data}
H. Friedrich.
\newblock Conformal structures of static vacuum data.
\newblock {\em Commun. Math. Phys.} 321 (2013) 419 - 482


\bibitem{friedrich:taylor-at-i}
H. Friedrich.
\newblock The Taylor expansion at  past time-like infinity.
\newblock {\em Commun. Math. Phys.} 324 (2013) 263 - 300. 

\bibitem{friedrich:ADS-stabiity}
H. Friedrich.
\newblock On the ADS stability problem.
\newblock {\em Class. Quantum Grav..} 31 (2014) 105001

\bibitem{friedrich:non-zero rest-mass}
H. Friedrich.
\newblock Smooth non-zero rest-mass evolution across time-like infinity.
\newblock    {\em Ann. Henri Poincare}  (2014), to appear.
\newblock arXiv:1311.0700 


\bibitem{friedrich:kannar1}
H. Friedrich, J. K\'ann\'ar.
\newblock Bondi systems near space-like infinity and the
calculation  of the NP-constants.
\newblock { J. Math. Phys.} 41, (2000), 2195 - 2232.


\bibitem{Geroch: 1977}
R. Geroch.
\newblock Asymptotic structure of space-time.
\newblock In: F. P. Esposito, L. Witten: {\it Asymptotic structure of space-time}.
\newblock Plenum, New York, 1977

\bibitem{Geroch-Horowitz:1978}
R. Geroch, G. T. Horowitz.
\newblock Asymptotically simple does not imply asymptotically Minkowskian.
\newblock {\it Phys. Rev. Lett.} 40 (1978) 203 - 206.

\bibitem{gibbons:hawking:1977}
G. W. Gibbons, S. W. Hawking.
\newblock Cosmological event horizons, thermodynamics, and particle creation.
\newblock {\it Phys. Rev.} D 15 (1977) 2738 - 2751.

\bibitem{graham:hirachi:2005}
R. Graham, K. Hirachi.
\newblock The ambient obstruction tensor and Q-curvature.
In: O. Biquard (ed.) {\it AdS/CFT Correspondence: Einstein metrics and their conformal boundaries}.
\newblock European Math. Soc., Z\"urich, 2005.


\bibitem{huebner:1996}
P. H\"ubner.
\newblock Method for calculating the global structure of (singular) space-times.
\newblock  {\it Phys. Rev.} D 53 (1996) 701 - 721.


\bibitem{huebner:2001}
P. H\"ubner.
\newblock From now to timelike infinity on a finite grid.
\newblock {\em Class. Quantum Grav.} 18 (2001) 1871 - 1884.


\bibitem{hawking:ellis:1973}
S. W. Hawking, G. F. R. Ellis.
\newblock {\it The large scale structure of space-time.}
\newblock Cambridge University Press, Cambridge, 1973.


\bibitem{kannar:1996a}
J. Kannar
\newblock On the existence of $C^{\infty}$ soutions to the asymptotic
characteristic initial value problem in general relativity.
\newblock {\em Proc. Roy. Soc.} A  452 (1996) 945 - 952.

\bibitem{kannar:1996b}
J. K\'ann\'ar.
\newblock Hyperboloidal initial data for the vacuum Einstein equations with cosmological constant.
\newblock {\em Class. Quantum Grav.} 13 (1996) 3075 - 3084.

\bibitem{kato:1975}
Kato, T. 
The Cauchy problem for quasi-linear symmetric hyperbolic systems,
{\it Arch. Ration. Mech. Anal.} 58 (1975)  181 - 205.

\bibitem{kichenassamy:2004}
S. Kichenassamy.
\newblock On a conjecture of Fefferman and Graham.
\newblock {\em Adv. in Math.} 184 (2004) 268 - 288.

\bibitem{komossa:2012}
S. Komossa.
\newblock Recoiling black holes: Electromagnetic signatures, candidates, and astrophysical implications.
\newblock Adv. Astronomy (2012) 364973.

\bibitem{kreiss:et:al:2009}
H.-O. Kreiss, O. Reula, O. Sarbach, J. Winicour.
\newblock Boundary conditions for coupled quasilinear wave equations with
application to isolated systems.
\newblock {\em Commun. Math. Phys.} 289 (2009) 1099 - 1129.

\bibitem{kundu:1981}
P. Kundu.
\newblock On the analyticity of stationary gravitational fields at spatial infinity.
\newblock {\it J. Math. Phys.} 22 (1981) 2006 - 2011.


\bibitem{luebbe:2013}
C. L\"ubbe.
\newblock Conformal scalar fields, isotropic singularities and conformal cyclic cosmologies.
\newblock  http://xxx.lanl.gov/abs/1312.2059

\bibitem{luebbe:valiente-kroon:2013}
C. L\"ubbe, J. A. Valiente Kroon.
\newblock A conformal approach to the analysis of the non-linear stability of radiation cosmologies.
\newblock {\it Ann. Phys.} 328 (2013) 1.

\bibitem{maldacena:1998}
J. M. Maldacena.
\newblock The large N limit of superconformal field theories and supergravity.
\newblock {\it Adv. Theor. Math. Phys.} 2 (1998) 231.






\bibitem{newman:penrose}
E. T. Newman, R. Penrose.
\newblock An approach to gravitational radiation by a methods of spin
coefficients.
\newblock{\em J. Math. Phys.} 3 (1962) 566 - 578.

\bibitem{newman:penrose:NP2}
E. T. Newman, R. Penrose.
\newblock New conservation laws for zero rest-mass fields in asymptotically
flat space-times.
\newblock{\em Proc. Roy. Soc. A} 305 (1968) 175 - 204.

\bibitem{RPAC-Newman: 1993}
R. P. A. C. Newman.
\newblock  On the structure of conformal singularities in classical general relativity. II Evolution equations and a conjecture of K. P. Tod.
\newblock {\it Proc. Roy. Soc.}  A 443 (1993) 493 - 515.

\bibitem{paetz:2013}
T.-T. Paetz.
\newblock Conformally covariant systems of wave equations and their equivalence to 
Einstein' s field equations.
\newblock arXiv: 1306.6204 [gr-qc]





\bibitem{penrose:1963} 
R. Penrose.
\newblock Asymptotic properties of fields and space-time.
\newblock {\em Phys. Rev. Lett.} 10 (1963) 66 - 68.

\bibitem{penrose:1965}
R. Penrose.
\newblock Zero rest-mass fields including gravitation: asymptotic behaviour.
\newblock {\it Proc. Roy. Soc. Lond.} A 284 (1965) 159 - 203.

\bibitem{penrose:1974}
R. Penrose. 
\newblock Relativistic Symmetry Groups.
\newblock In: A. O. Barut (ed.): {\it Group theory in non-linear problems},
\newblock Reidel, Dordrecht, 1974

\bibitem{penrose:2010}
R. Penrose.
\newblock Cycles of Time.
\newblock Bodley Head, 2010.


\bibitem{pirani:1957}
F. A. E. Pirani.
\newblock Invariant Formulation of Gravitational Radiation Theory.
\newblock {\em Phys. Rev.} 105 (1957) 1089 - 1099. 



\bibitem{rendall:2004}
A. Rendall.
\newblock Asymptotics of solutions of the Einstein equations with positive cosmological constant.
\newblock   {\em Ann. Henri Poincar\'e}  5 (2004) 1041 - 1064.

\bibitem{ringstrom:2008}
H. Ringstr\"om.
\newblock  Future stability on the Einstein non-linear scalar field system.
\newblock   {\em Invent. math.}  173 (2008) 123 -  208.


\bibitem{rinne:moncrief:2013}
O. Rinne, V. Moncrief.
\newblock Hyperboloidal Einstein-matter evolution and tails for scalar and Yang-Mills fields.
\newblock {\em Class. Quantum Grav.} 30 (2013) 095009




\bibitem{sachs:waves VI}
R. K. Sachs.
\newblock Gravitational waves in general relativity VI. 
The outgoing radiation condition.
\newblock { Proc. Roy. Soc} A 264 (1961) 309--338.

\bibitem{sachs:waves VIII}
R. K. Sachs.
\newblock Gravitational waves in general relativity VIII. Waves in
asymptotically flat space-time.
\newblock { Proc. Roy. Soc} A 270 (1962) 103--126.


\bibitem{starobinsky:1983}
A. A. Starobinsky.
\newblock Isotropization of arbitrary cosmological expansion given an effective cosmological constant.
\newblock   {\em JETP Lett.} 37 (1983) 66 - 69


\bibitem{tod:1987}
K. P. Tod.
\newblock Quasi-local mass and cosmological singularities.
\newblock {\it Class. Quantum Grav.} 4 (1987) 1457 - 1468.



\bibitem{tod:2003}
K. P. Tod.
\newblock Isotropic cosmological singularities: other matter models.
\newblock {\it Class. Quantum Grav.} 20 (2003) 521 - 534.

\bibitem{tod:2007}
K. P. Tod.
\newblock Isotropic cosmological singularities in spatially homogeneous models with cosmological constant.
\newblock {\it Class. Quantum Grav.} 24 (2007) 2415 - 2432.

\bibitem{trautman:1958a}
A. Trautman.
\newblock Lectures on General Relativity.
\newblock Reprinted in {\it Gen. Rel. Grav.} 34 (2002) 721 - 762.



\bibitem{trautman:1958c}
A. Trautman.
\newblock Radiation and boundary conditions in the theory of gravitation.
\newblock {\em Bull. Acad. Pol. Sci., S\'erie sci. math., astr. et phys.} VI
(1958) 407 - 412.


\bibitem{j.a.v.kroon:2004a}
J. A. Valiente Kroon.
\newblock A new class of obstructions to the smoothness of null infinity.
\newblock {\it Commun. Maht. Phys.} 244 (2004) 133 - 156.


\bibitem{j.a.v.kroon:2007}
J. A. Valiente Kroon.
\newblock Asymptotic properties of the development of
conformally flat data near spatial infinity.
\newblock {\it Class. Quantum Grav.} 24 (2007) 3037 - 3053




\end{thebibliography}
\end{document}